\documentclass[prd,twocolumn,showpacs,amsmath,amssymb,nofootinbib,superscriptaddress]{revtex4}
\usepackage{graphicx}
\usepackage[pdftex]{color}
\usepackage{bm}
\usepackage{amsfonts}
\usepackage[latin1]{inputenc}
\usepackage{amssymb}
\usepackage{color}
\usepackage{float}
\usepackage{amsmath}
\usepackage{dcolumn}
\usepackage{hyperref}
\usepackage{amsfonts}

\newcommand{\data}{\ensuremath{{\mathcal D}}}
\newcommand{\model}{\ensuremath{{\mathcal M}}}
\newcommand{\erf}{\ensuremath{{\rm{erf}}}}
\newcommand{\nuu}{\ensuremath{{\tilde{\nu}}}}

\def\nn{\nonumber}

\def\be{\begin{equation}}
\def\ee{\end{equation}}
\def\bea{\begin{eqnarray}}
\def\eea{\end{eqnarray}}
\def\eqi{\begin{equation}}
\def\eqf{\end{equation}}
\def\eqia{\begin{eqnarray}}
\def\eqfa{\end{eqnarray}}

\begin{document}

\title{Testing Einstein's gravity and dark energy with growth of matter perturbations: Indications for new Physics?}

\author{Spyros Basilakos}\email{svasil@academyofathens.gr}
\affiliation{Academy of Athens, Research Center for Astronomy and Applied Mathematics, Soranou Efesiou 4, 11527, Athens, Greece}

\author{Savvas Nesseris}\email{savvas.nesseris@csic.es}
\affiliation{Instituto de F\'isica Te\'orica UAM-CSIC, Universidad Auton\'oma de Madrid, Cantoblanco, 28049 Madrid, Spain}

\date{\today}

\begin{abstract}
The growth index of matter fluctuations is computed for ten distinct accelerating cosmological models and confronted to the latest growth rate data via a two-step process. First, we implement a joint statistical analysis in order to place constraints on the free parameters of all models using solely background data. Second, using the observed growth rate of clustering from various galaxy surveys we test the performance of the current cosmological models at the perturbation level while either marginalizing over $\sigma_8$ or having it as a free parameter. As a result, we find that at a statistical level, i.e. after considering the best-fit $\chi^2$ or the value of the Akaike information criterion, most models are in very good agreement with the growth rate data and are practically indistinguishable from $\Lambda$CDM. However, when we also consider the internal consistency of the models by comparing the theoretically predicted values of $(\gamma_0, \gamma_1)$, i.e. the value of the growth index $\gamma(z)$ and its derivative today, with the best-fit ones, we find that the predictions of three out of ten dark energy models are in mild tension with the best-fit ones when $\sigma_8$ is marginalized over. When $\sigma_8$ is free we find that most models are not only in mild tension, but also predict low values for $\sigma_8$. This could be attributed to either a systematic problem with the growth-rate data or the emergence of new physics at low redshifts, with the latter possibly being related to the well-known issue of the lack of power at small scales. Finally, by utilizing mock data based on an LSST-like survey we show that with future surveys and by using the growth index parameterization, it will be possible to resolve the issue of the low $\sigma_8$ but also the tension between the fitted and theoretically predicted values of $(\gamma_0, \gamma_1)$.
\end{abstract}

\maketitle

\section{Introduction}
The majority of studies in observational cosmology converge to the following general conclusion (see Refs.\cite{Hicken2009,Ade:2015xua} and references therein), that the Universe is spatially flat and it contains $\sim 30\%$ of matter (luminous and dark), while the rest is the enigmatic dark energy (DE). Despite the great progress made at theoretical level, up to now the nature of the DE has yet to be discovered and several unanswered questions remain, see Refs.\cite{Perivolaropoulos:2011hp}, \cite{Perivolaropoulos:2014lua} for an overview and a discussion of some of the problems. As a matter of fact, the discovery of the underlying physics of dark energy, thought to be driving the accelerated expansion of the Universe, is considered one of the most fundamental problems on the interface uniting astronomy, cosmology and particle physics.

In the literature there is a large family of cosmological scenarios that provide a mathematical explanation regarding the accelerated expansion of the Universe. Generally speaking, the cosmological models are mainly classified in two large categories. The first group of DE models is nested inside Einstein's general relativity (GR) and it introduces new fields in nature (for review see \cite{Ame10} and references therein). Alternatively, modified gravity models provide a theoretical platform which assumes that the present accelerating epoch is due to the possibility of gravity becoming weak at extragalactic scales. Therefore, DE has nothing to do with new fields and it appears as a geometric effect \cite{Ame10}.

In this framework, the corresponding effective equation-of-state (EoS) parameter is allowed to take values in the phantom regime, namely $w<-1$ (for other possible explanations see \cite{Cai10} and \cite{BB14}). Notice, that the cosmological implications of modified gravity models have been reviewed in the article of Clifton et al. \cite{Clif12}. Of course, one may also choose to follow a model-independent approach to reconstruct the expansion history of the Universe, eg see Refs.~\cite{Nesseris:2010ep} and \cite{Nesseris:2012tt}.

On the other hand, the growth of matter perturbations is a key test for studies of matter distribution in the Universe \cite{Oku2015}, and, more importantly, it can readily be accessed from observations. Specifically, the growth rate of clustering has been measured based on galaxy surveys, like SDSS, BOSS, {\em WiggleZ} etc (see our Table \ref{tab:fsigma8data} and references therein). The growth of matter perturbations can be also be used for self-consistency tests of general relativity, see Refs.~\cite{Nesseris:2014mfa},\cite{Nesseris:2014qca}, even in a model independent fashion.

However, from the theoretical viewpoint, it has been proposed that the so-called growth index $\gamma$, first introduced by \cite{Peeb93}, can be used towards testing the nature of dark energy. Indeed, in the literature one can find a large body of studies in which the theoretical form of the growth index is provided analytically for various cosmological models, including scalar field DE \cite{Silv94,Wang98,Linjen03,Lue04,Linder2007,Nes08}, DGP \cite{Linder2007,Gong10,Wei08,Fu09}, $f(R)$ \cite{Gann09,Tsu09}, Finsler-Randers \cite{Bastav13}, time varying vacuum models $\Lambda(H)$, \cite{Basola}), Clustered DE \cite{Mehra2015a}, Holographic dark energy \cite{Mehra2015} and $f(T)$ \cite{BasFT}.

In this article, we attempt to check the performance of a large family of flat DE models (10 models) at the perturbation level. First, a joint likelihood analysis, involving the latest geometrical data (SNe type Ia, CMB shift parameter and BAO) is performed in order to  determine the cosmological parameters of the DE models. Second, we attempt to discriminate the different DE cosmologies by estimating the growth index $\gamma$ and the corresponding redshift evolution. Then, by utilizing the available growth rate data we show that the evolution of the growth index is a potential discriminator for a large fraction of the explored DE models.

The structure of the manuscript is as follows: In Sec. II we present the main ingredients of the linear growth of matter fluctuations in the dark energy regime. In Secs. III and IV with the aid of a joint statistical analysis (based on SNe Ia, CMB shift parameter and BAO data) we constrain the DE model parameters. In Sec. V, we test the DE cosmologies by comparing the corresponding theoretical predictions of the growth index evolution with observations. Finally, we summarize our conclusions in Sec. VI.

\section{Linear growth and dark energy}
In this section we provide the basic tools that are necessary in order to study the linear matter fluctuations. Since we are well inside in the matter dominated era we can neglect the radiation term from the Hubble expansion. Now, for different types of dark energy the differential equation that governs the evolution of matter fluctuations at subhorizon scales is \cite{Lue04,Linder2007,Gann09,Stab06,Uzan07,Tsu08,Steigerwald:2014ava}
\begin{equation}
\label{eq:111}
\ddot{\delta}_{m}+2\nuu H\dot{\delta}_{m}-4\pi G\mu \rho_{m} \delta_{m}=0 \;.
\end{equation}
As is well known, $\delta_{m} \propto D(t)$, where $D(t)$ is the linear growth factor usually normalized to unity at the present time. It is clear that the nature of dark energy is reflected in the quantities $\nuu$ and $\mu\equiv G_{\rm eff}/G_{N}$\footnote{However, in more complicated models, eg ones with couplings between matter and geometry, one may have $\mu\equiv G_{\rm eff}/G_{N}+\beta(a,k)$, where $\beta(a,k)$ is a function that depends on derivatives of the Lagrangian of the model \cite{Nesseris:2008mq}.}. In the case of scalar field dark energy models which
adhere to Einstein's gravity we have $\nuu=\mu=1$, while if we allow interactions in the dark sector, in general we get $\nuu \neq 1$ and $\mu\neq 1$. For either inhomogeneous dark energy models (inside GR) or modified gravity models one can show that $\nuu=1$ and $\mu\ne 1$.

Another important quantity in this kind of study is the growth rate of clustering (first introduced by \cite{Peeb93})
\begin{equation}
\label{fzz221}
f(a)=\frac{d\ln \delta_{m}}{d\ln a}\simeq \Omega^{\gamma}_{m}(a)\;.
\end{equation}
Based on the above equation we can easily obtain the growth factor
\be
\label{eq244}
D(a)={\rm exp} \left[\int_{1}^{a}
\frac{\Omega_{m}(x)^{\gamma(x)}}{x} dx \right]\;,
\ee
with
\begin{equation}
\label{ddomm}
\Omega_{m}(a)=\frac{\Omega_{m0}a^{-3}}{E^{2}(a)}
\end{equation}
and from which we define
\begin{equation}
\label{ddomm1}
\frac{d\Omega_{m}}{da}=-3\frac{\Omega_{m}(a)}{a}\left( 1+\frac{2}{3}
\frac{d{\rm ln}E}{d{\rm ln}a} \right) \;.
\end{equation}
Notice, that $E(a)=H(a)/H_{0}$ is the dimensionless Hubble parameter and $\gamma$ is the so called growth index. Therefore, inserting Eq.(\ref{fzz221}) in
Eq.(\ref{eq:111}) and with the aid of Eq.(\ref{ddomm1}) we arrive at
\be \label{fzz444}
a\frac{df}{da}+
\left(2\nuu+\frac{d{\ln}E}{d{\rm ln}a}\right)f+f^{2}
=\frac{3\mu \Omega_{m}}{2}
\ee
or
{\small{
\begin{equation}
\label{Poll}
a{\rm ln}(\Omega_{m})\frac{d\gamma}{da}+\Omega_{m}^{\gamma}
-3\gamma+2\nuu-
\left(\gamma-\frac{1}{2}\right)\frac{d{\ln}E}{d{\rm ln}a}=\frac{3}{2}
\mu\Omega_ { m } ^ { 1-\gamma}.
\end{equation}}}
Another expression of the above equation is given by
Steigerwald et al. \cite{Steigerwald:2014ava}
\be
\frac{d\omega}{d{\ln}a}(\gamma+\omega\frac{d\gamma}{d\omega}) +{\rm
e}^{\omega \gamma}+2\nuu+\frac{d{\ln}E}{d{\rm
ln}a}=\frac{3}{2}\mu {\rm e}^{\omega(1-\gamma)},
\ee
where $\omega={\rm ln}\Omega_{m}(a)$ which means that at $z\gg 1$ ($a \to 0$) we have
$\Omega_{m}(a)\to 1$ [or $\omega \to 0$]. In this context, Steigerwald et al. \cite{Steigerwald:2014ava}
found a useful formula which provides the asymptotic value of the growth index (see Eq.(8) in \cite{Steigerwald:2014ava} and the relevant discussion in \cite{Basola})
\be
\label{g000}
\gamma_{\infty}=\frac{3(M_{0}+M_{1})-2(H_{1}+N_{1})}{2+2X_{1}+3M_{0}},
\ee
where the relevant quantities are
\be \label{Coef1}
M_{0}=\left. \mu \right|_{\omega=0}\,,
\ \
M_{1}=\left.\frac{d \mu}{d\omega}\right|_{\omega=0}
\ee
and
\be \label{Coef2}
N_{1}=\left.\frac{d \nuu}{d\omega}\right|_{\omega=0}\,,\ \
H_{1}=-\frac{X_{1}}{2}=\left.\frac{d \left(d{\rm ln}E/d{\rm ln}a\right)}{d\omega}\right|_{\omega=0} \,.
\ee

Concerning the functional form of the growth index we use a Taylor expansion around $a(z)=1$
(see \cite{Pol,Bel12,DP11,Ishak09,Bass})
\be
\label{aPoll1}
\gamma(a)=\gamma_{0}+\gamma_{1}(1-a).
\ee
Therefore, the asymptotic value reduces to $\gamma_{\infty}\simeq \gamma_{0}+\gamma_{1}$, where we have set
$\gamma_{0}=\gamma(1)$. Now, writing Eq.(\ref{Poll}) at the present epoch ($a=1$)
\begin{eqnarray}
\label{Poll1}
\hspace{-0.4cm}&&-\gamma^{\prime}(1){\rm
ln}(\Omega_{m0})+\Omega_{m0}^{\gamma(1)}-3\gamma(1)+2\nuu_{0}
-2(\gamma_{0}-\frac{1}{2})\left.\frac{d{\ln}E}{d{\rm ln}a}\right|_{a=1}\nonumber\\
\hspace{-0.4cm}&& =\frac {3}{2} \mu_ { 0 } \Omega_{m0}^{1-\gamma(1)}, \ \ \
\end{eqnarray}
and using Eq.(\ref{aPoll1}) we find
\begin{equation}
\label{Poll2}
\gamma_{1}=\frac{\Omega_{m0}^{\gamma_{0}}-3\gamma_{0}+2\nuu_{0}
-2(\gamma_{0}-\frac{1}{2})\left.\frac{d{\ln}E}{d{\rm ln}a}\right|_{a=1}
-\frac{3}{2}\mu_{0}\Omega_{m0}^{1-\gamma_{0}} }
{\ln  \Omega_{m0}}\;.
\end{equation}
Notice, that a prime denotes a derivative with respect to the scale factor, $\mu_{0}=\mu(1)$ and $\nuu_{0}=\nuu(1)$.

To conclude this section it is important to realize that the growth of matter perturbations is affected by the main cosmological functions, namely $E(a)$, $\Omega_{m}(a)$, $\mu(a)$ and $\nuu(a)$. Therefore, for the benefit of the reader let us briefly present the main steps that we follow in the rest of the paper.

\begin{itemize}
\item Suppose that we have a dark energy model that contains $n$-free cosmological parameters, given by the cosmological vector $\theta^{i}=(\theta^{1},\theta^{2},...,\theta^{n})$. First we place constraints on $\theta^{i}$ by performing an overall likelihood analysis, involving the latest geometrical data (standard candles and standard rulers).

\item For this cosmological model we know its basic cosmological quantities which implies that we can compute $\gamma_{\infty}$ from Eq.(\ref{g000}). Then solving the system of $\gamma_{\infty}=\gamma_{0}+\gamma_{1}$ and Eq.(\ref{Poll2}) we can write $\gamma_{0,1}$ in terms of the cosmological parameters $\theta^{i}$ ($\Omega_{m0}$, etc).

\item Once, steps (i) and (ii) are accomplished, we finally test the performance of the cosmological model at the perturbation level utilizing the available growth data.
\end{itemize}

\section{Likelihood Analysis}
In this section we perform a joint statistical analysis using the latest background data. Briefly, the total
likelihood function is the product of the individual likelihoods:
\begin{equation}\label{eq:like-tot}
 {\cal L}_{\rm tot}(\theta^{i})={\cal L}_{\rm sn} \times {\cal L}_{\rm bao} \times {\cal L}_{\rm cmb}\;,
\end{equation}
thus the overall chi-square $\chi^2_{\rm tot}$ is written as
\begin{equation}\label{eq:like-tot_chi}
 \chi^2_{\rm tot}(\theta^{i})=\chi^2_{\rm sn}+\chi^2_{\rm bao}+\chi^2_{\rm cmb}.
\end{equation}
In particular we use the JLA SNIa data of Ref.~\cite{Betoule:2014frx}, the BAO from 6dFGS\cite{Beutler:2011hx}, SDDS\cite{Anderson:2013zyy}, BOSS CMASS\cite{Xu:2012hg}, WiggleZ\cite{Blake:2012pj}, MGS\cite{Ross:2014qpa} and BOSS DR12\cite{Gil-Marin:2015nqa}. Finally, we also use the CMB shift parameters based on the \textit{Planck 2015} release \cite{Ade:2015xua}, as derived in Ref.~\cite{Wang:2015tua}.

As we have already mentioned in the previous section, the cosmological vector $\theta^{i}$ includes the free parameters of the particular cosmological model which are related with the cosmic expansion. In the present analysis, some of the relevant parameters are
$\theta^{i}=(\alpha,\beta,\Omega_{m0},\Omega_{d0},\Omega_{r0},\Omega_{b0}, H_0,...)$, where $\Omega_{m0}$ and $\Omega_{b0}$ are the total matter (cold dark matter and baryons)
and baryon density parameters today, while $\alpha, \beta$ are the parameters related to the stretch and color of the SNIa data. Assuming a spatially flat universe we have, $\Omega_{d0}=1-\Omega_{m0}-\Omega_{r0}$. Also, in the case of the CMB shift parameter, the contribution of the radiation term $\Omega_{r0}$ and the baryon density $\Omega_{b0}$ needs to be considered (see below). Here the radiation density at the present epoch is fixed to $\Omega_{r0}=\Omega_{m0} a_{eq}$, where the scale factor at equality is $a_{eq}=\frac{1}{1+2.5~10^4 \Omega_{m0} h^2 \left(T_{cmb}/2.7K\right)^{-4}}$.

We also marginalize over the parameters $M$ and $\delta M$ of the JLA set as described in the appendix of Ref.~\cite{Conley:2011ku}. These parameters implicitly contain $H_0$ and thus $\chi^2_{\rm sn}$ is independent of $H_{0}=h~100km/s/Mpc$, where according to Planck $h\simeq0.67$ \cite{Ade:2015xua}. However, we keep the parameters $\alpha, \beta$ free in our analysis. Therefore, using the aforementioned arguments the cosmological vector becomes $\theta^{i}=(\alpha, \beta,\Omega_{m0},\Omega_{b0}h^2, h,\theta^{i+1}_{d})$, where $\theta^{i+1}_{d}$ contains the free parameters which are related with the nature of the dark energy.

The next step is to apply the Akaike Information Criterion (AIC) information criterion \cite{Akaike:1974} in order to
test the statistical performance of the models themselves. Since, $N/k \gg 1$ the AIC formula is given by
$$
{\rm AIC} = -2 \ln {\cal L}_{\rm max}+2k\;,
$$
where ${\cal L}_{\rm max}$ is the maximum likelihood, $N$ is the number of data points used in the fit and $k$ is the number of free parameters. A smaller value of AIC indicates a better model-data fit. In the case of Gaussian errors, $\chi^{2}_{min}=-2{\cal L}_{\rm max}$, one can show that the difference in AIC between two models is written as $\Delta {\rm AIC}=\Delta \chi^{2}_{\rm min}+2\Delta k$.

\section{Constraints on Dark Energy Models}\label{sec.constr}
Here we provide the basic properties of the most popular DE models whose free parameters are constrained following the methodology of the previous section. We mention that in all cases we assume a spatially flat Friedmann-Lema\^\i tre-Robertson-Walker (FLRW) geometry.

Notice, that for the study of matter perturbations in Secs. II and V the effect of radiation is not necessary. However, for the fitting of the current DE models to the Baryonic Acoustic Oscillations (BAO) and the CMB shift parameter in Sec. III we need to include the radiation component in the Hubble parameter. In order to deal with this issue we replace the matter component $\Omega_{m0}a^{-3}$ in the normalized Hubble parameter $E(a)$ with $\Omega_{m0}a^{-3}+\Omega_{r0}a^{-4}$. Accordingly, the present value of $\Omega_{d0}=1-\Omega_{m0}$ is replaced by $\Omega_{d0}=1-\Omega_{m0}-\Omega_{r0}$.

In Table \ref{tab:background}, the reader may see a more compact presentation of the best fit values of cosmological parameters $\theta^{i}$, including also the various nuisance parameters $\alpha,\beta$ of the JLA SnIa data, the separate contribution of the baryons $\Omega_{b0}h^2$ but also the best fit $\chi^2_{min}$ and the corresponding value of the AIC. In what follows we will focus on $\Omega_{m0}$ and the various DE model parameters, but for completeness in Table \ref{tab:background} we give the best-fit values of the other parameters ($\alpha, \beta,\Omega_{b0}h^2, h$) as well.

\subsection{Constant equation of state ($w$CDM model)}
In this simple model the equation of state (hereafter EoS) parameter $w=p_{d}/\rho_{d}$ is constant \cite{MT97}, where $p_{d}$ is the pressure and $\rho_{d}$ is the density of the dark energy fluid respectively. Although the quintessence scenario ($-1\le w<-1/3$), driven by a real scalar field, suffers from the extreme fine tuning it has been widely used in the literature due to its simplicity. On the other hand, we remind the reader that the DE models which obey $w<-1$ are endowed with exotic physics, namely a scalar field with a negative kinetic term, usually called phantom dark energy \cite{phantom}.

The $w$CDM model adheres to GR and it does not allow interactions in the dark sector, namely $\mu(a)=\nuu(a)=1$. Also, the dimensionless Hubble parameter is given by
\be
\label{EWCDM}
E^{2}(a)=\Omega_{m0}a^{-3}+\Omega_{d0}a^{-3(1+w)},
\ee
where $\Omega_{d0}=1-\Omega_{m0}$. Therefore, from Eq.(\ref{EWCDM}) we arrive
at
\be
\label{EWCDM11}
\frac{d{\ln}E}{d{\rm ln}a}=
-\frac{3}{2}-\frac{3}{2}w\left[1-\Omega_{m}(a)\right]
\ee
and
$$
\{ M_{0},M_{1},H_{1},X_{1}\}=\{ 1,0,\frac{3w}{2},-3w \}\;.
$$
If we substitute the above coefficients into Eq.(\ref{g000}) then
we find (see also
\cite{Silv94,Wang98,Linjen03,Lue04,Linder2007,Nes08,Bass})
$$
\gamma_{\infty} = \frac{3(w-1)}{6w-5}.
$$
Note however, that the above expression neglects the effects of DE perturbations, as discussed in Ref.~\cite{Nesseris:2015fqa}. Of course for $w=-1$ we fully recover the $\Lambda$CDM model in which $\gamma_{\infty}^{(\Lambda)} = 6/11$.

From a statistical point of view, the cosmological vector reduces to $\theta^{i}=(\Omega_{m0},w)$. In this case the total likelihood function peaks at $\Omega_{m0}=0.320\pm 0.004$ and $w=-0.983\pm 0.012$ with
$\chi_{\rm min}^{2}(\Omega_{m},w) \simeq 708.438$ (AIC=720.438). The final step is to estimate the pair
$(\gamma_{0},\gamma_{1})$. Based on the procedure described at the end of section III and
utilizing the best fit values of the cosmological parameters (see Table \ref{tab:background}) we obtain $(\gamma_{0},\gamma_{1}) \simeq (0.557,-0.011)$. Lastly, considering the concordance $\Lambda$ cosmology
and minimizing with respect to $\Omega_{m0}$ we find $\Omega_{m0}=0.317\pm 0.003$ with
$\chi^{2}_{\rm min}(\Omega_{m0}) \simeq 708.592$ (AIC=718.592) and $(\gamma_{0},\gamma_{1}) \simeq (0.556,-0.011)$.

We should note that the $\Lambda$CDM value for $\Omega_{m0}$ is in excellent agreement with the one obtained from the Planck 2015 TT, TE, EE and lowP CMB data $\Omega_{m0}^{Planck}=0.3156\pm 0.0091$ \cite{Ade:2015xua}, which confirms that our analysis with the CMB shift parameters correctly captures the expansion history of the Universe as measured by Planck\footnote{Note that the frequently quoted value of $\Omega_{m0}^{Planck}=0.308\pm 0.012$, e.g. see the Abstract of the Planck paper \cite{Ade:2015xua}, besides the TT,TE,EE and lowP CMB data also includes the lensing trispectrum data, a piece of information which is not captured by the CMB shift parameters. By solely comparing the CMB data however, we can see that our result is in excellent agreement as described in the text.}.

\subsection{Parametric dark energy (CPL model)}
This kind of phenomenological model was first introduced by Chevalier-Polarski-Linder \cite{Chevallier:2001qy,Linder:2002et}. In particular, the dark energy EoS parameter is parameterized
as a first order Taylor expansion around the present epoch:
$$
w(a)=w_{0}+w_{1}(1-a),\label{cpldef}
$$
where $w_{0}$ and $w_{1}$ are constants, while for an interesting extension of this model see Ref.~\cite{Pantazis:2016nky}. The normalized Hubble parameter now becomes:
$$
E^{2}(a)=\Omega_{m0}a^{-3}+\Omega_{d0}
a^{-3(1+w_{0}+w_{1})}e^{3w_{1}(a-1)}.
$$
Since the CPL model is inside GR and due to the absence of dark matter/energy interactions we get $\mu(a)=\nuu(a)=1$. Also the logarithmic derivative $d{\rm ln}E/d{\rm ln}a$ is given by Eq.(\ref{EWCDM11}) but here we have $w=w(a)$. Using the above functions we can derive the growth coefficients (see also \cite{Steigerwald:2014ava})
$$
\{ M_{0},M_{1},H_{1},X_{1}\}=\{ 1,0,\frac{3(w_{0}+w_{1})}{2},-3(w_{0}+w_{1}) \}\;,
$$
which provide
$$
\gamma_{\infty} = \frac{3(w_{0}+w_{1}-1)}{6(w_{0}+w_{1})-5} .
$$
In this case the cosmological vector contains three free parameters $\theta^{i}=(\Omega_{m0},w_{0},w_{1})$ and
the overall likelihood function peaks at $\Omega_{m0}=0.320 \pm 0.005$, $w_{0}=-1.007 \pm 0.0001$ and $w_{1}=0.111\pm 0.009$. The corresponding $\chi_{\rm min}^{2}(\Omega_{m0},w_{0},w_{1})$ is 708.283 (AIC=722.283)
and $(\gamma_{0},\gamma_{1}) \simeq (0.556,-0.008)$.

\subsection{HDE model}
Applying the holographic \cite{Hola} principle within the framework of GR $\nuu(a)=1$ one can show that
$$
w(a)=-\frac{1}{3}-\frac{2\sqrt{\Omega_{\rm d}(a)}}{3s}
$$
and
$$
\frac{d{\rm ln}\Omega_{d}}{d{\rm ln}a}=-\frac{w(a)}{3}\left[ 1-\Omega_{d}(a)\right],
$$
where $\Omega_{d}(a)=1-\Omega_{m}(a)$  and $s$ is a constant.
It is easy to check that at high redshifts
$z\gg 1$ ($a \to 0$ and $\Omega_{d} \to 0$)
the asymptotic value of the EoS parameter $w_{\infty}$ tends to
$-1/3$. Also
the dimensionless Hubble parameter, $E(a)=H(a)/H_0$ is given by
$$
E^2(z)=\frac{\Omega_{m0}a^{-3}}{1-\Omega_{d}(a)}\;.
$$
Again, the functional form of $d{\rm ln}E/d{\rm ln}a$ is given by Eq.(\ref{EWCDM}).
Obviously, the above three equations produce a system whose solution gives the evolution of the main cosmological parameters, namely $E(a)$, $w(a)$ and $\Omega_{d}(a)$, where
$\theta^{i}=(\Omega_{m0},s)$.

The quantity $\mu(a)$ that describes the intrinsic features of the HDE is written as \cite{Mehra2015}
\begin{equation} \label{VV}
\mu(a)=\left\{ \begin{array}{cc} 1
\;\;
       &\mbox{homogeneous HDE}\\
  1+\frac{\Omega_{\rm d}(a)}{\Omega_{\rm m}(a)}\Delta_{\rm d}(a)(1+3c_{\rm eff}^2)
\;\;
       & \mbox{clustered HDE}
       \end{array}
        \right.
\end{equation}
where $\Delta_{d}=\frac{1+w(a)}{1-3w(a)}$ \cite{Bat,Mehra2015} and $c_{\rm eff}^{2}$ is the effective sound speed of the dark energy.

Here we consider the following two cases:
\begin{itemize}

\item Homogeneous HDE (hereafter HHDE) in which $\mu(a)=1$. Therefore, in this case we find (see \cite{Mehra2015})
$$
\{ M_{0},M_{1},H_{1},X_{1}\}=\{ 1,0,\frac{3w_{\infty}}{2},-3w_{\infty} \}\;.
$$
where $w_{\infty}\simeq -1/3$ and the asymptotic value of the
growth index is
$$
\gamma_{\infty} = \frac{4}{7} \;.
$$
Our joint statistical analysis yields that the
likelihood function peaks at $\Omega_{m0}=0.311\pm 0.003$ and $s=0.654\pm
0.006$ with $\chi_{\rm min}^{2}(\Omega_{m0},s) \simeq 713.218$ (AIC=725.218).
Using the latter cosmological parameters we estimate
$(\gamma_{0},\gamma_{1}) \simeq (0.558,0.013)$.

\item Clustered HDE (hereafter CHDE): here $\mu(a)$ is given by the second branch of Eq.(\ref{VV}). In this framework, we obtain
$$
\{ M_{0},M_{1},H_{1},X_{1}\}=\{ 1,-\frac{(1+3c_{\rm eff}^2)}{3},\frac{3{\rm w}_{\infty}}{2},-3{\rm w}_{\infty}\}
$$
and for $w_{\infty}=-1/3$ we get via Eq.(\ref{g000})
$$
\gamma_{\infty}=\frac{3(1-c_{\rm eff}^2)}{7} \;.
$$
We restrict our analysis to $c_{\rm eff}^{2}=0$, which implies that the sound horizon is small with respect to the Hubble radius and thus DE perturbations grow in a similar fashion to matter perturbations \cite{Arm}. Utilizing the aforementioned cosmological parameters we compute $(\gamma_{0},\gamma_{1}) \simeq (0.541,-0.112)$. For more details concerning the analytical derivation of the growth index we refer the reader to \cite{Mehra2015}.

\end{itemize}

\subsection{Time varying vacuum ($\Lambda_{t}$CDM model)}
Let us now consider the possibility of a decaying $\Lambda$-cosmology, that is, $\Lambda=\Lambda(a)$. The decaying vacuum equation of state does not depend on whether $\Lambda$ is strictly constant or variable. Therefore, the EoS takes the nominal form, $p_{\Lambda}(t)=-\rho_{\Lambda}(t)=-\Lambda(t)/8\pi G$. In the current article we study a specific dynamical vacuum model which is based on the renormalization group in quantum field theory. It has been proposed that, the evolution of the vacuum is given by
$$
\Lambda(H)=\Lambda_0+ 3\nu\,(H^{2}-H_0^2)\,,
$$
where $\Lambda_0\equiv\Lambda(H_0)=3\Omega_{\Lambda 0}H^{2}_{0}$ and $\nu$ is provided in the Renormalization Group (RG) context as a ``$\beta$-function which determines the running of the cosmological ``constant'' (CC) within QFT in curved spacetime\,\cite{Sola2013}. The Friedmann equations are the same with those of the concordance $\Lambda$CDM model, while the current vacuum scenario matter is obliged to exchange energy with vacuum in order to fulfil the Bianchi identity which implies
$$
\dot{\rho}_{m}+3H\rho_{m}=-\dot{\rho}_{\Lambda}.
$$
Combining the Friedmann equations and the latter generalized conservation law one can write the evolution of the normalized Hubble parameter\footnote{For the fitting the radiation term is included in the $\Lambda_{t}$CDM model as follows: we replace ${\tilde \Omega_{m0}}a^{-3(1-\nu)}$ in Eq.(\ref{anorm11}) with
${\tilde \Omega_{m0}}a^{-3(1-\nu)}+{\tilde \Omega_{r0}}a^{-4(1-\nu)}$, where
${\tilde \Omega_{m 0}}\equiv \frac{\Omega_{m0}}{1-\nu}$,
${\tilde \Omega_{r 0}}\equiv \frac{\Omega_{r0}}{1-\nu}$,
and
${\tilde \Omega_{\Lambda 0}}\equiv \frac{1-\Omega_{m0}-\Omega_{r0}-\nu}{1-\nu}$
\cite{GoSolBas2015}.}
\begin{equation}
\label{anorm11}
E^{2}(a)=
{\tilde \Omega}_{\Lambda 0}+{\tilde \Omega_{m0}}a^{-3(1-\nu)}\;,
\end{equation}
with
\be
\label{EE1}
\frac{d{\ln}E}{d{\rm ln}a}=
-\frac{3}{2}(1-\nu){\tilde \Omega}_{m}(a)\;,
\ee
where we have set
${\tilde \Omega}_{m}(a)=
\frac{\tilde \Omega_{m0}a^{-3(1-\nu)}}{E^{2}(a)}$,
${\tilde \Omega_{m 0}}\equiv \frac{\Omega_{m0}}{1-\nu}$ and
${\tilde \Omega_{\Lambda 0}}\equiv \frac{1-\Omega_{m 0}-\nu}{1-\nu}$.
Obviously, the cosmological vector includes the following free parameters $\theta^{i}=(\alpha,\beta,{\tilde \Omega_{m0}},\Omega_{b0}h^2,h,\nu)$. For more details concerning the global dynamics of the present time varying vacuum model we refer the reader to the following Refs.\cite{BPS09,Grande11,BasSol2014,GoSolBas2015,GomezSola2015,Basrev}.

On the other hand, the growth index of matter perturbations has been investigated by Basilakos and Sola \cite{Basola}. Specifically, the quantities $\nuu$ and $\mu$ are given by
\be
\label{nn1}
\nuu=1+\frac32\,\nu
\ee
and
\be
\label{mm1}
\mu(a)=
1-\nu-\frac{4\nu}{{\tilde \Omega}_{m}(a)}+3\nu(1-\nu).
\ee
To this end, the growth coefficients are found to be \cite{Basola}
$$
\{ M_{0},M_{1},H_{1},X_{1}\}= \{ 1-2\nu-3\nu^{2},
-\frac{3(1-\nu)}{2},3(1-\nu) \}
$$
which provide
$$
\gamma_{\infty}=
\frac{6+3\nu}{11-12\nu}\;.
$$
Finally, using the cosmological data and the joint likelihood analysis we find
$(\Omega_{m},\nu)=(0.320\pm0.002,-1.41\cdot10^{-4}\pm 2.91\cdot10^{-4})$ for a best-fit
$\chi^{2}_{\rm min}(\Omega_{m0},\nu)\simeq 708.55$ (AIC=720.550).
Based on the aforementioned cosmological parameters we obtain
$(\gamma_{0},\gamma_{1})=(0.572,-0.023)$.

\subsection{Dvali, Gabadadze and Porrati (DGP) gravity}
The first modified gravity model that we present is that of Dvali, Gabadadze and Porrati \cite{Dvali2000}. In this scenario, one can obtain an accelerating expansion of the Universe based on the fact that gravity itself becomes weak at cosmological scales (close to Hubble radius) because our four dimensional spacetime survives into an extra dimensional manifold (see \cite{Dvali2000} and references therein). It has been shown that the normalized Hubble parameter is written as
\be
\label{eos222g}
E(a)=\sqrt{\Omega_{m0} a^{-3}+\Omega_{rc}}+\sqrt{\Omega_{rc}},
\ee
where
$\Omega_{rc}=(1-\Omega_{m0})^{2}/4$ for a matter only Universe\footnote{When we also include radiation, this changes to $\Omega_{rc}=(1-\Omega_{m0}-\Omega_{r0})^{2}/4$}.
From Eq.(\ref{eos222g}), we easily find
\be
\label{eos222e}
\frac{d{\rm ln}E}{d{\rm ln}a}=
-\frac{3\Omega_{m}(a)}{1+\Omega_{m}(a)}.
\ee
For the DGP model the function $\mu(a)$ takes the form
$$
\mu(a)=\frac{2+4\Omega^{2}_{m}(a)}{3+3\Omega^{2}_{m}(a)}
$$
and $\nuu(a)=1$. Inserting the above equations in Eqs.(\ref{Coef1}),
(\ref{Coef2}) we have
$$
\{ M_{0},M_{1},H_{1},X_{1}\}=\{ 1,\frac{1}{3},-\frac{3}{4},\frac{3}{2}\}
$$
and from Eq.(\ref{g000}) the asymptotic value of the growth index
becomes (see also \cite{Linder2007,Gong10,Wei08,Fu09})
$$
\gamma_{\infty}=\frac{11}{16}.
$$

As in the concordance $\Lambda$CDM model, the cosmological vector contains the following free parameters $\theta^{i}=(\alpha,\beta,\Omega_{m0},\Omega_{b0}h^2,h)$. The overall statistical analysis provides a best fit value of $\Omega_{m}=0.392\pm
0.008$, but the fit is much worse, $\chi_{\rm min}^{2}(\Omega_{m})\simeq 798.654$ (AIC=808.654), with respect to that of $\Lambda$CDM cosmology. To this end, using the above and the calculations of section III we obtain
$(\gamma_{0},\gamma_{1})\simeq (0.669,0.019)$.

\subsection{Finsler-Randers dark energy model (FRDE)}
In the last decade there have been quite interesting applications of Finsler geometry in its Finsler-Randers version, in the topics of cosmology, astrophysics and general relativity \cite{Stav07} (and references therein). Recently, it has been found \cite{Bastav13} that the Finsler-Randers field equations provide a Hubble parameter which is identical with that of DGP gravity. This means that Eqs.(\ref{eos222g}) and (\ref{eos222e}) are also valid here. As expected, the joint analysis provides exactly the same statistical results.

However, the two models (FRDE and DGP) deviate at the perturbation level since in the case of the FRDE model we have $\mu(a)={\tilde \nu}(a)=1$ \cite{Stav07}. Therefore, it is easy to show that
$$
\{ M_{0},M_{1},H_{1},X_{1}\}=\{ 1,0,-\frac{3}{4},\frac{3}{2}\}
$$

$$
\gamma_{\infty}= \frac{9}{16}.
$$
Again, solving the system $\gamma_{\infty}=\gamma_{0}+ \gamma_{1}$ and Eq.(\ref{Poll2}) for $\Omega_{m}=0.392\pm 0.008$ we derive $(\gamma_{0},\gamma_{1})\simeq (0.566,-0.003)$.

\subsection{Power law $f(T)$ gravity model}
Among the large family of modified gravity models, the $f(T)$ gravity extends the old definition of the so called teleparallel equivalent of general relativity \cite{ein28,Hayashi79,Maluf:1994ji}, where $T$ is the
torsion scalar. In the current article we use the power-law model of Bengochea and Ferraro \cite{Ben09}, with
$$
f(T)=\alpha (-T)^{b},
$$
where
$$
\alpha=(6H_0^2)^{1-b}\frac{\Omega_{F0}}{2b-1} \;.
$$
In this framework, the Hubble parameter normalized to unity at the present time takes the form
\begin{eqnarray}
\label{Mod1Ezz}
E^2(a,b)=\Omega_{m0}a^{-3}+\Omega_{d0} E^{2b}(a,b)
\;,
\end{eqnarray}
where $\Omega_{d0}=1-\Omega_{m0}$. Obviously, for $b=0$ the power law $f(T)$ model reduces to $\Lambda$CDM cosmology\footnote{Notice, that for $b=1/2$ it reduces to the Dvali, Gabadadze and Porrati (DGP)
ones \cite{Dvali2000}.}, namely $T+f(T)=T-2\Lambda$ (where $\Lambda=3\Omega_{d0}H_{0}^{2}$, $\Omega_{d0}=\Omega_{\Lambda 0}$). It has been shown that in order to have an accelerated expansion of the Universe which is consistent with the cosmological data one needs $b\ll 1$ \cite{Linder:2010py,Nesseris2013}. Therefore, following the notations of Nesseris et al. \cite{Nesseris2013} we perform a first order Taylor expansion of $E^2(a,b)$ around $b=0$ and thus we obtain an approximate normalized Hubble parameter, namely
\be
\label{approxM1}
E^2(a,b)\simeq E^2_\Lambda(a)+\Omega_{d0}\ln\left[E^2_\Lambda(a)\right]b+... \;.
\ee

\begin{table*}[t!]
\centering
\caption{A summary of the best-fit background parameters for the various cosmological models used in the analysis. The fifth and sixth columns show the specific DE model parameters.}
\begin{tabular}{cccccccccc}
\hline
\hline
Model & $\alpha$  & $\beta$ & $\Omega_{m0}$ & $\Omega_{b0}h^2$ & h& DE Params  & $\chi^2_{min}$ & AIC \\
\hline
$\Lambda$CDM  & $0.141\pm0.004$ & $3.097\pm0.011$ & $0.317\pm0.003$ & $0.0222\pm0.0001$ & $0.672\pm0.003$ & $w_0=-1$,~$w_a=0$ &  $708.592$ & $718.592$\\
\hline
$w$CDM  & $0.141\pm0.006$ & $3.097\pm0.010$ & $0.320\pm0.004$ & $0.0222\pm0.0001$ & $0.668\pm0.004$ & $w_0=-0.983\pm0.012$ &  $708.438$ & $720.438$\\
& & & & & & $w_a=0$ &  & \\
\hline
CPL  & $0.141\pm0.004$ & $3.083\pm0.012$ & $0.320\pm0.005$ & $0.0223\pm0.0002$ & $0.667\pm0.004$ & $w_0=-1.008\pm0.009$ &  $708.283$ & $722.283$\\
  &  &  &  &  &  & $w_a=0.111\pm0.066$   &  & \\
\hline
HDE  & $0.142\pm0.004$ & $3.122\pm0.010$ & $0.311\pm0.003$ & $0.0224\pm0.0001$ & $0.673\pm0.003$ & $s=0.654\pm0.006$ &  $713.218$ & $725.218$\\
\hline
$\Lambda_t$CDM  & $0.141\pm0.005$ & $3.113\pm0.004$ & $0.320\pm0.002$ & $0.0222\pm0.0001$ & $0.672\pm0.004$ & $\nu=(-1.4\pm2.9)\cdot10^{-4}$ &   $708.550$ & $720.550$\\
\hline
DGP-FRDE  & $0.136\pm0.004$ & $3.079\pm0.010$ & $0.392\pm0.008$ & $0.0229\pm0.0015$ & $0.587\pm0.004$ &    & $798.654$ & $808.654$\\
\hline
$f(T)$  & $0.141\pm0.006$ & $3.095\pm0.015$ & $0.320\pm0.004$ & $0.0222\pm0.0001$ & $0.667\pm0.003$ & $b=0.038\pm0.008$ &   $708.363$ & $720.363$\\
\hline
$f(R)$  & $0.141\pm0.009$ & $3.099\pm0.020$ & $0.319\pm0.007$ & $0.0222\pm0.0002$ & $0.670\pm0.005$ & $b=0.091\pm0.009$ &   $708.526$ & $720.526$\\
\hline
\hline
\end{tabular}
\label{tab:background}
\end{table*}

Recently, Basilakos \cite{BasFT} investigated the growth index for the power law $f(T)$ gravity model. In brief, differentiating Eq.(\ref{Mod1Ezz}) and using Eq.(\ref{ddomm}) we arrive at
\be
\frac{d{\ln}E}{d{\rm ln}a}=
-\frac{3}{2}\frac{\Omega_{m}(a)}{[1-bE^{2(b-1)}\Omega_{d0}]}
\ee
and for $b\ll 1$ we find
\be
\label{Taylor2}
\frac{d{\ln}E}{d{\rm ln}a}\simeq
-\frac{3}{2}\Omega_{m}(a)\left[ 1+\frac{\Omega_{d0}b}{E^{2}_{\Lambda}(a)}+...\right].
\ee
Notice that here we have ${\tilde \nu}=0$ and the quantity $\mu$ takes the following form (see \cite{BasFT} and references therein)
\begin{equation}
\mu(a)=\frac{1}{1+ \frac{b\Omega_{d0}} {(1-2b)E^{2(1-b)}}}
\end{equation}
or
\be
\label{Geff1}
\mu(a)\simeq 1-\frac{\Omega_{d0}}{E^{2}_{\Lambda}(a)}\;b+ \cdots .
\ee
Now, based on the above equations the growth index coefficients of
(\ref{Coef1}) and (\ref{Coef2}) become (for more details see \cite{BasFT})
$$
\{ M_{0},M_{1},H_{1},X_{1}\}= \{ 1,b,-\frac{3(1-b)}{2},3(1-b)\}
$$
and thus the asymptotic value of the growth index is
$$
\gamma_{\infty} = \frac{6}{11-6b} \;.
$$
As expected, for $b=0$ we recover the $\Lambda$CDM value $6/11$. In this context, we find that the likelihood function peaks at $\Omega_{m0}=0.320\pm 0.004$, $b=0.038\pm 0.008$ with $\chi^{2}_{\rm min}(\Omega_{m0},b)\simeq 708.363$ (AIC=720.363). Therefore, if we use the latter best fit solution then we estimate
$(\gamma_{0},\gamma_{1}) \simeq (0.564,-0.007)$.

\subsection{$f(R)$ gravity ($f$CDM model)}
Another modified gravity that we include in our analysis is the popular $f(R)$ model of Hu and Sawicki. However, here we make use of the implementation with the $b$ parameter as in Ref.~\cite{Basilakos:2013nfa}. This has two advantages: first, the deviation from $\Lambda$CDM is easily seen and second, by doing a series expansion around $b=0$ we can find extremely accurate (better than $0.1\%$ for $b\lesssim1$ and better than $10^{-5}\%$ for $b\lesssim0.1$) analytical approximations. In this formalism, the Lagrangian for the Hu and Sawicki model can be equivalently written as \cite{Basilakos:2013nfa}:
\be
\label{Hu1}
f(R)= R- \frac{2\Lambda }{1+\left(\frac{b \Lambda }{R}\right)^n}
\ee
where $n$ is a parameter of the model, henceforth chosen as $n=1$ without loss of generality.

As mentioned we can perform a series expansion of the solution of the equations on motion around $b=0$, i.e. $\Lambda$CDM, as
\be
H^2(a)=H_{\Lambda}^2(a)+\sum_{i=1}^M b^i \delta H_i^2(a), \label{expansion1}
\ee
where
\be
\frac{H_{\Lambda}^2(a)}{H_0^2}=\Omega_{m0}a^{-3}+\Omega_{r0} a^{-4}+(1-\Omega_{m0}-\Omega_{r0})\label{LCDM1}
\ee and $M$ is the number of terms we keep before truncating the series, but usually only the two first non-zero terms are more than enough for excellent agreement with the numerical solution. Finally, $\delta H_i^2(a)$ is a set of algebraic functions that can be determined from the equations of motion, see Ref.~\cite{Basilakos:2013nfa} for the exact and quite long expressions.

Studying the growth index in this class of models is more complicated, as the modified Newton's constant depends on both the time via the scale factor $a$ and the scale $k$, ie $G_{\rm eff}=G_{\rm eff}(a,k)$ \cite{Tsujikawa:2007gd}. More specifically we have
\be
\frac{G_{\rm eff}(a,k)}{G_N}=\frac{1}{F}\frac{1+4\frac{k^2}{a^2}F_{,R}/F}{1+3\frac{k^2}{a^2}F_{,R}/F},\label{geff}
\ee
where $F=f'(R)$, $F_{,R}=f''(R)$, $G_N$ is the bare Newton's constant and we have normalized Eq.~(\ref{geff}) so that for $b=0$, i.e. for the $\Lambda$CDM we get $\frac{G_{\rm eff}(a,k)}{G_N}=1$ as expected. We also follow Ref.~\cite{Basilakos:2013nfa} and set $k=0.1h {\rm Mpc}^{-1}\simeq 300 H_0$. In the notation of the other sections we have:
\bea
\mu(a,k)&=&\frac{G_{\rm eff}(a,k)}{G_N},\nn \\
\tilde{\nu}(a)&=&1.
\eea

In Ref.~\cite{Gannouji:2008wt} it was shown that these kinds of models predict rather low and rather high values for the parameters $\gamma_0$ and $\gamma_1$ respectively or more specifically $(\gamma_0,\gamma_1)\simeq(0.4,-0.2)$. Because of the $k$-dependence of the effective Newton's constant in order to get the exact values for these parameters we need to solve Eq.~(\ref{fzz444}) numerically to estimate $\gamma_0\simeq\frac{\ln(f(1))}{\ln(\Omega_{m0})}$, where $f(1)$ is the growth rate at $a=1$, and then use Eq.~(\ref{Poll2}) to get $\gamma_1$.

In this case the cosmological vector contains two free parameters $\theta^{i}=(\Omega_{m0},b)$ and the overall likelihood function peaks at $\Omega_{m0}=0.319 \pm 0.007$ and $b=0.091 \pm 0.009$. The corresponding $\chi_{\rm min}^{2}(\Omega_{m0},b)$ is 708.526 (AIC=720.526) and $(\gamma_{0},\gamma_{1}) \simeq (0.395,-0.294)$.

\begin{table}[t!]
\centering
\caption{The $f\sigma_8(z)$ growth data.}
\begin{tabular}{ccc}
\hline
\hline
z & $f\sigma_8(z)$ & Ref. \\
\hline
$0.02$  & $0.360\pm0.040$ & \cite{Hudson:2012gt}\\
$0.067$ & $0.423\pm0.055$ & \cite{Beutler:2012px}\\
$0.10$  & $0.370\pm0.130$   & \cite{Feix:2015dla}\\
$0.17$  & $0.510\pm0.060$ & \cite{Percival:2004fs}\\
$0.35$  & $0.440\pm0.050$ & \cite{Song:2008qt,Tegmark:2006az}\\
$0.77$  & $0.490\pm0.180$ & \cite{Guzzo:2008ac,Song:2008qt}\\
$0.25$  & $0.351\pm0.058$ & \cite{Samushia:2011cs}\\
$0.37$  & $0.460\pm0.038$ & \cite{Samushia:2011cs}\\
$0.22$  & $0.420\pm0.070$ & \cite{Blake:2011rj}\\
$0.41$  & $0.450\pm0.040$ & \cite{Blake:2011rj}\\
$0.60$  & $0.430\pm0.040$ & \cite{Blake:2011rj}\\
$0.60$  & $0.433\pm0.067$ & \cite{Tojeiro:2012rp}\\
$0.78$  & $0.380\pm0.040$ & \cite{Blake:2011rj}\\
$0.57$  & $0.427\pm0.066$ & \cite{Reid:2012sw}\\
$0.30$  & $0.407\pm0.055$ & \cite{Tojeiro:2012rp}\\
$0.40$  & $0.419\pm0.041$ & \cite{Tojeiro:2012rp}\\
$0.50$  & $0.427\pm0.043$ & \cite{Tojeiro:2012rp}\\
$0.80$  & $0.470\pm0.080$ & \cite{delaTorre:2013rpa}\\
\hline
\hline
\end{tabular}
\label{tab:fsigma8data}
\end{table}

\section{Testing dark energy models with growth data}
\subsection{Analysis with the real data}
In this section we present the details of the statistical method and on the observational sample that we adopt in order to test the performance of the dark energy models at the perturbation level. Specifically, we utilize the recent growth rate data, namely $A\equiv f(z)\sigma_{8}(z)$ where $\sigma_{8}(z)$ is the redshift-dependent rms fluctuations of the linear density field at at $R=8h^{-1}$Mpc. Notice that the sample contains $N_{\rm gr}=18$ entries (see Table \ref{tab:fsigma8data} and the corresponding references). Following the standard analysis we use the $\chi^2$-minimization procedure, which in our case is defined as follows:
\be
\label{Likel}
\chi^{2}_{\rm gr}(\phi^{\mu})=
\sum_{i=1}^{N_{\rm gr}} \left[ \frac{A_{\cal D}(z_{i})-
A_{\cal M}(z_{i},{\bf \phi^{\mu}})}
{\sigma_{i}}\right]^{2}
\ee
where $\phi^{\mu} =(\sigma_{8}\equiv \phi^{1},\gamma_{0},\gamma_{1})$ is the statistical vector at the background level (not to be confused with $\theta^{i}$), $A_{\cal D}(z_{i})$ and $\sigma^{2}_{i}$ are the growth data and the corresponding uncertainties at the observed redshift $z_{i}$\footnote{In the case of $\Lambda_{t}$CDM model we remind the reader that we need to replace $\Omega_{m}(z)$ with ${\tilde \Omega}_{m}(z)={\tilde \Omega}_{m0}(1+z)^{3}/E^{2}(z)$}. Also, ${\cal D}$ and ${\cal M}$ indicate data and model respectively. The theoretical growth-rate is given by:
\be
\label{Likell}
A_{\cal M}(z,\phi^{\mu})=f\sigma_{8}(z,\phi^{\mu})=
\sigma_{8}D(z)\Omega_{m}(z)^{\gamma(z)}\;.
\ee
where for the latter equality we have set $\sigma_{8}(z)=\sigma_{8}D(z)$, $\Omega_{m}(z)=\Omega_{m0}(1+z)^{3}/E^{2}(z)$, $D(z)$ is the growth factor normalized to unity at the present
time and $\gamma(z)$ is the growth index.

Now, we provide the basic steps towards marginalizing $\chi^{2}_{gr}$ over $\sigma_{8}$ (see also \cite{Tadd2015}). Substituting the second equality Eq.(\ref{Likell}) into Eq.(\ref{Likel}), we simply obtain
\begin{equation}\label{eq:expand-xi2}
\chi^2_{\rm gr}=\Gamma-2B\sigma_8+C\sigma_{8}^2\;,
\end{equation}
where
\begin{eqnarray}\label{eq:xi2-A-B-C}
 \Gamma &=& \sum\limits_{i=1}^{N_{\rm gr}}\frac{A^{2}_{\cal D}(z_{i})}{\sigma_{i}^2}\;,\nonumber \\
 B &=&\sum\limits_{i=1}^{N_{\rm gr}}\frac{A_{\cal D}(z_{i})D(z_{i})\Omega_{m}(z_{i})^{\gamma}}
{\sigma_{i}^2}\;,\nonumber \\
 C &=& \sum\limits_{i=1}^{N_{\rm gr}} \frac{D^{2}(z_{i})\Omega_{m}(z_{i})^{2\gamma}}{\sigma_{i}^{2}}\;. \nonumber
\end{eqnarray}
The corresponding likelihood $L_{\rm gr}= e^{-\chi^{2}_{\rm gr}/2}$ is then given by
\begin{equation}
L_{\rm gr}(\data|\phi^\mu,\model)  = e^{ -\frac{1}{2}\left[\Gamma  -\frac{B^2}{C}
  + C \left( \frac{B}{C} -  \sigma_8 \right)^2 \right]},
\end{equation}
where we have completed the square. Applying Bayes's theorem and marginalizing over $\sigma_8$ we find
\bea
 \hspace{-0.4cm} p(\phi^{\mu}| \data,\model) &=&  \frac{1}{p(\data|\model)}  e^{ -\frac{1}{2} \left[\Gamma  -\frac{B^2}{C} \right]}\cdot \nn \\ && \int d\sigma_8  \; p(\sigma_8, \phi^{\mu}|\model) e^{ -\frac{C}{2}  \left( \frac{B}{C} -\sigma_8 \right)^2 }.
\eea

Considering flat priors, namely $p(\sigma_8, \phi^{\mu}|\model) = 1$
and $\sigma_8$ is within a range $\sigma_8\in ( 0, \infty ) $ we arrive at
\begin{equation}
 p(\phi^{\mu}| \data,\model) =  \frac{1}{p(\data|\model)}  e^{ -\frac{1}{2} \left[\Gamma  -\frac{B^2}{C} \right]}
\int_0^\infty d\sigma_8  e^{ -\frac{C}{2}  \left( \frac{B}{C} -  \sigma_8 \right)^2 } \;.
\end{equation}
Introducing the variable $ y =  \sigma_8 - \frac{B}{C}$ we find
\begin{equation}
 p(\phi^{\mu}| \data,\model) =
\frac{1}{p(\data|\model)}  e^{ -\frac{1}{2} \left[\Gamma  -\frac{B^2}{C} \right]}
\sqrt{\frac{\pi}{2C}} \left[ 1 +  \erf \left(  \frac{B}{\sqrt{2C} }\right) \right],
\end{equation}
where $\erf(x) = \int_0^x dy e^{-y^2}$,
to which corresponds the marginalized ${\tilde \chi}^2_{\rm gr}$ function
\begin{equation}
\label{eq:marg_chi2}
{\tilde \chi}^{2}_{\rm gr}=
\Gamma  -\frac{B^2}{C}
+  \ln C
- 2 \ln \left[ 1 +  \erf \left(  \frac{B}{\sqrt{2C} }\right) \right]\;.
\end{equation}
Notice that we have ignored the constant $-  \ln \frac{\pi}{2}$. The first two terms in ${\tilde \chi}^{2}_{\rm gr}$, i.e. $\Gamma  -\frac{B^2}{C}$ correspond to the case where $\sigma_8$ is fixed in such a way that the original $\chi^{2}_{\rm gr}$ [see Eq.(\ref{Likel})] is minimized.

In what follows we consider two approaches: first, we use the marginalized ${\tilde \chi}^2_{\rm gr}$ function which is independent of $\sigma_8$ and thus it contains only two free parameters $(\gamma_{0},\gamma_{1})$ and second, for the sake of comparison we also of minimize $\chi^{2}_{\rm gr}$, given by Eq.~(\ref{Likel}), with respect to $\sigma_8$.

Lastly, in order to compare the DE models we utilize the Akaike information criterion for small sample size which is defined for the case of Gaussian errors, as:
$$
{\rm AIC}={\tilde \chi}^2_{\rm gr, min}+2k_{\rm gr}+\frac{2k_{\rm gr}(k_{\rm gr}-1)}{N_{\rm gr}-k_{\rm gr}-1},
$$
where $k_{\rm gr}$ is the number of free parameters.

\begin{center}
\begin{table}
\caption{A summary of the best-fit parameters $(\gamma_0,\gamma_1)$ for the various cosmological models used in the analysis, with $\sigma_8$ marginalized over and the $\chi^2$ given by Eq.~(\ref{eq:marg_chi2}).}
\begin{tabular}{ccccc}
\hline
\hline
Model & $\gamma_0$  & $\gamma_1$ & $\chi^2_{min}$ & AIC  \\
\hline
$\Lambda$CDM  & $0.556\pm0.170$ & $-0.438\pm1.415$ & $13.456$ & $17.723$\\
\hline
$w$CDM  & $0.561\pm0.173$ & $-0.453\pm1.426$ & $13.454$ & $17.721$\\
\hline
CPL     & $0.564\pm0.176$ & $-0.430\pm1.454$ & $13.445$ & $17.712$\\
\hline
HDE     & $0.565\pm0.177$ & $-0.235\pm1.592$ & $13.412$ & $17.679$\\
\hline
$\Lambda_t$CDM   & $0.560\pm0.171$ & $-0.438\pm1.426$ & $13.455$ & $17.722$\\
\hline
DGP-FRDE & $0.714\pm0.341$ & $-0.714\pm1.931$ & $13.409$ & $17.676$\\
\hline
$f(T)$   & $0.562\pm0.175$ & $-0.448\pm1.436$ & $13.451$ & $17.718$\\
\hline
$f(R)$   & $0.567\pm0.176$ & $-0.453\pm1.447$ & $13.450$ & $17.717$\\
\hline
\hline
\end{tabular}
\label{tab:growth}
\end{table}
\end{center}

\begin{center}
\begin{table}
\hspace{-.1cm}\caption{A summary of the best-fit parameters $(\gamma_0,\gamma_1,\sigma_8)$ for the various cosmological models used in the analysis and the $\chi^2$ given by Eq.~(\ref{Likel}).}
\begin{tabular}{cccccc}
\hline
\hline
\hspace{-0.20cm}Model & $\gamma_0$  & $\gamma_1$ & $\sigma_8$ & $\chi^2_{min}$ & AIC  \\
\hline
\hspace{-0.20cm}$\Lambda$CDM  & $0.545\pm0.122$ & $-0.558\pm0.532$ & $0.687\pm0.095$ & $6.999$ & $13.856$\\
\hline
\hspace{-0.20cm}$w$CDM  & $0.550\pm0.122$ & $-0.576\pm0.532$ & $0.687\pm0.097$ & $6.998$ & $13.855$\\
\hline
\hspace{-0.20cm}CPL     & $0.552\pm0.122$ & $-0.558\pm0.533$ & $0.689\pm0.099$ & $6.995$ & $13.852$\\
\hline
\hspace{-0.20cm}HDE     & $0.553\pm0.122$ & $-0.382\pm0.543$ & $0.699\pm0.104$ & $6.992$ & $13.849$\\
\hline
\hspace{-0.20cm}$\Lambda_t$CDM   & $0.549\pm0.123$ & $-0.559\pm0.537$ & $0.687\pm0.095$ & $6.999$ & $13.856$\\
\hline
\hspace{-0.20cm}DGP & $0.670\pm0.143$ & $-0.992\pm0.555$ & $0.687\pm0.154$ & $6.985$ & $13.842$\\
\hspace{-0.20cm}FRDE &  &  &  &  & \\
\hline
\hspace{-0.20cm}$f(T)$   & $0.551\pm0.122$ & $-0.573\pm0.532$ & $0.688\pm0.098$ & $6.998$ & $13.855$\\
\hline
\hspace{-0.20cm}$f(R)$   & $0.555\pm0.123$ & $-0.580\pm0.535$ & $0.688\pm0.098$ & $6.997$ & $13.854$\\
\hline
\hline
\end{tabular}
\label{tab:growth1}
\end{table}
\end{center}

Below and in Table \ref{tab:growth}, we provide our statistical results for the case when $\sigma_8$ is marginalized over.

\begin{itemize}
  \item For the $w$CDM model:
${\tilde \chi}^2_{\rm gr, min}=13.454$ (AIC=17.221),
$\gamma_{0}=0.561\pm  0.173$ and $\gamma_{1}=-0.453\pm 1.426$.
Regarding the $\Lambda$CDM cosmological model our best fit solution
is $\gamma_{0}=0.556\pm  0.170$ and $\gamma_{1}=-0.438\pm 1.415$ with
${\tilde \chi}^2_{\rm gr, min}=13.456$ (AIC=17.723), which is agreement with
that of \cite{Nesseris2013,Pour};

  \item For the CPL model: ${\tilde \chi}^2_{\rm gr, min}
=13.445$ (AIC=17.712),
$\gamma_{0}=0.564\pm  0.176$ and $\gamma_{1}=-0.430\pm 1.454$.

  \item For the HDE model: ${\tilde \chi}^2_{\rm gr, min}
=13.412$ (AIC=17.679),
$\gamma_{0}=0.565\pm  0.177$ and $\gamma_{1}=-0.235\pm 1.592$.

  \item For the $\Lambda_{t}$CDM model: ${\tilde \chi}^2_{\rm gr, min}
=13.455$ (AIC=17.722),
$\gamma_{0}=0.567\pm  0.066$ and $\gamma_{1}=0.116\pm 0.191$;

  \item For the DGP-FRDE gravity: ${\tilde \chi}^2_{\rm gr, min}
=13.409$ (AIC=17.676),
$\gamma_{0}=0.714\pm  0.341$ and $\gamma_{1}=-0.714\pm 1.931$;

  \item For the $f(T)$ model: ${\tilde \chi}^2_{\rm gr, min}
=13.451$ (AIC=17.718),
$\gamma_{0}=0.562\pm  0.175$ and $\gamma_{1}=-0.448\pm 1.436$;

  \item For the $f$CDM model: ${\tilde \chi}^2_{\rm gr, min}
=13.450$ (AIC=17.717),
$\gamma_0 = 0.567\pm 0.176$ and $\gamma_1
= -0.453\pm 1.447$.
\end{itemize}

Below and in Table \ref{tab:growth1}, we also provide our statistical results for the case when $\sigma_8$ is a free parameter.

\begin{itemize}
  \item For the $w$CDM model:
$\chi^2_{\rm gr, min}=6.998$ (AIC=13.855),
$\gamma_{0}=0.550\pm  0.122$, $\gamma_{1}=-0.576\pm 0.532$ and $\sigma_8=0.687\pm0.097$.
Regarding the $\Lambda$CDM cosmological model our best fit solution
is $\gamma_{0}=0.545\pm  0.122$, $\gamma_{1}=-0.558\pm 0.532$ and $\sigma_8=0.687\pm0.095$ with
$\chi^2_{\rm gr, min}=6.999$ (AIC=13.856), which is agreement with
that of \cite{Nesseris2013,Pour};

  \item For the CPL model: $\chi^2_{\rm gr, min}
=6.995$ (AIC=13.852),
$\gamma_{0}=0.552\pm  0.122$, $\gamma_{1}=-0.558\pm 0.533$ and $\sigma_8=0.689\pm0.099$.

  \item For the HDE model: $\chi^2_{\rm gr, min}
=6.992$ (AIC=13.849),
$\gamma_{0}=0.553\pm  0.122$, $\gamma_{1}=-0.382\pm 0.543$ and $\sigma_8=0.699\pm0.104$.

  \item For the $\Lambda_{t}$CDM model: $\chi^2_{\rm gr, min}
=6.999$ (AIC=13.856),
$\gamma_{0}=0.549\pm  0.123$, $\gamma_{1}=-0.559\pm 0.537$ and $\sigma_8=0.689\pm0.095$.

  \item For the DGP-FRDE gravity: $\chi^2_{\rm gr, min}
=6.985$ (AIC=13.842),
$\gamma_{0}=0.670\pm  0.143$,d $\gamma_{1}=-0.992\pm 0.555$ and $\sigma_8=0.687\pm0.154$.

  \item For the $f(T)$ model: $\chi^2_{\rm gr, min}
=6.998$ (AIC=13.855),
$\gamma_{0}=0.551\pm  0.122$, $\gamma_{1}=-0.573\pm 0.532$ and $\sigma_8=0.688\pm0.098$.

  \item For the $f$CDM model: $\chi^2_{\rm gr, min}
=6.997$ (AIC=13.854),
$\gamma_0 = 0.555\pm 0.123$, $\gamma_1= -0.580\pm 0.535$ and $\sigma_8=0.688\pm0.098$.
\end{itemize}

\begin{figure*}[!t]
\centering
\includegraphics[width = 0.245\textwidth]{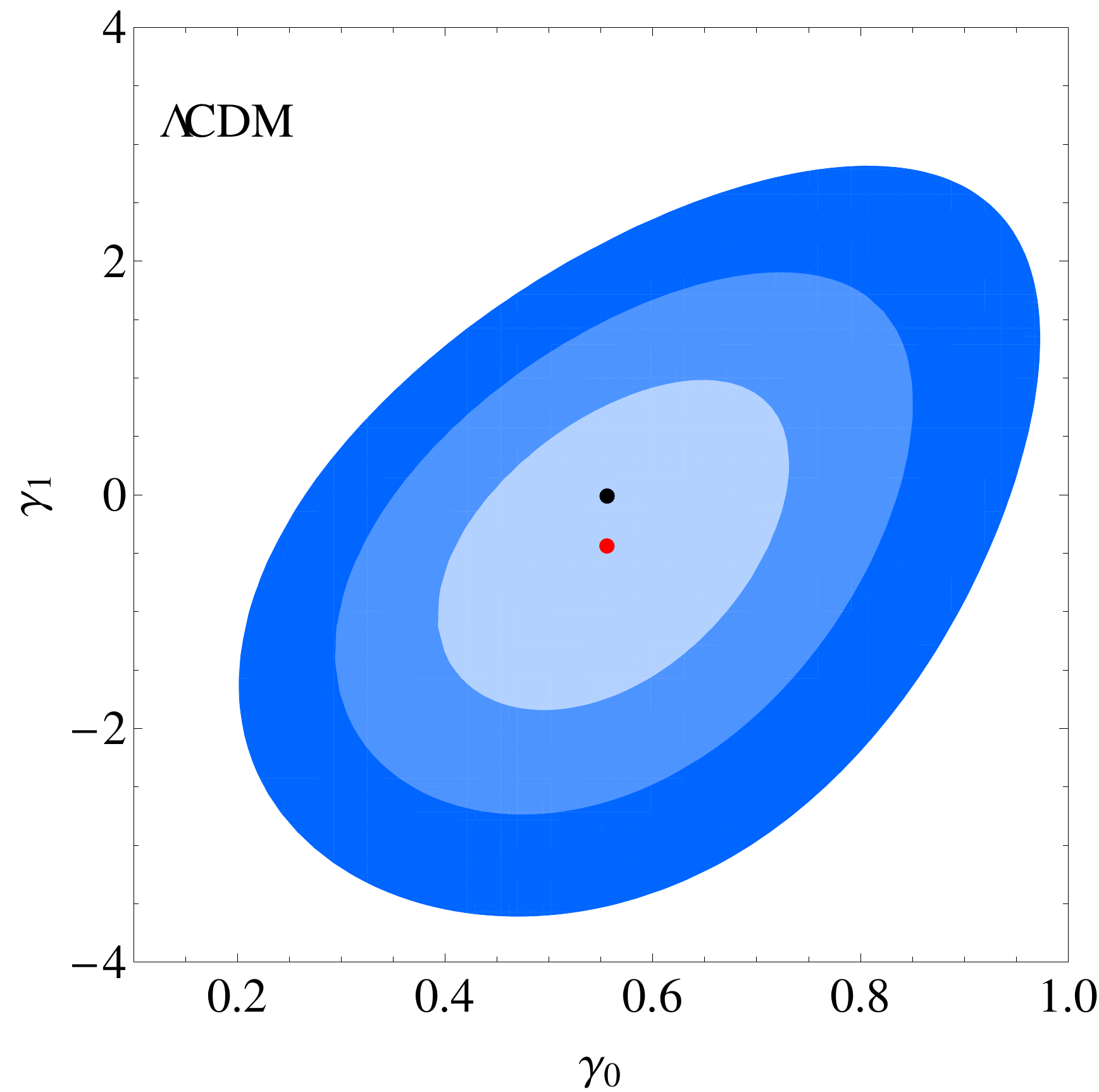}
\includegraphics[width = 0.245\textwidth]{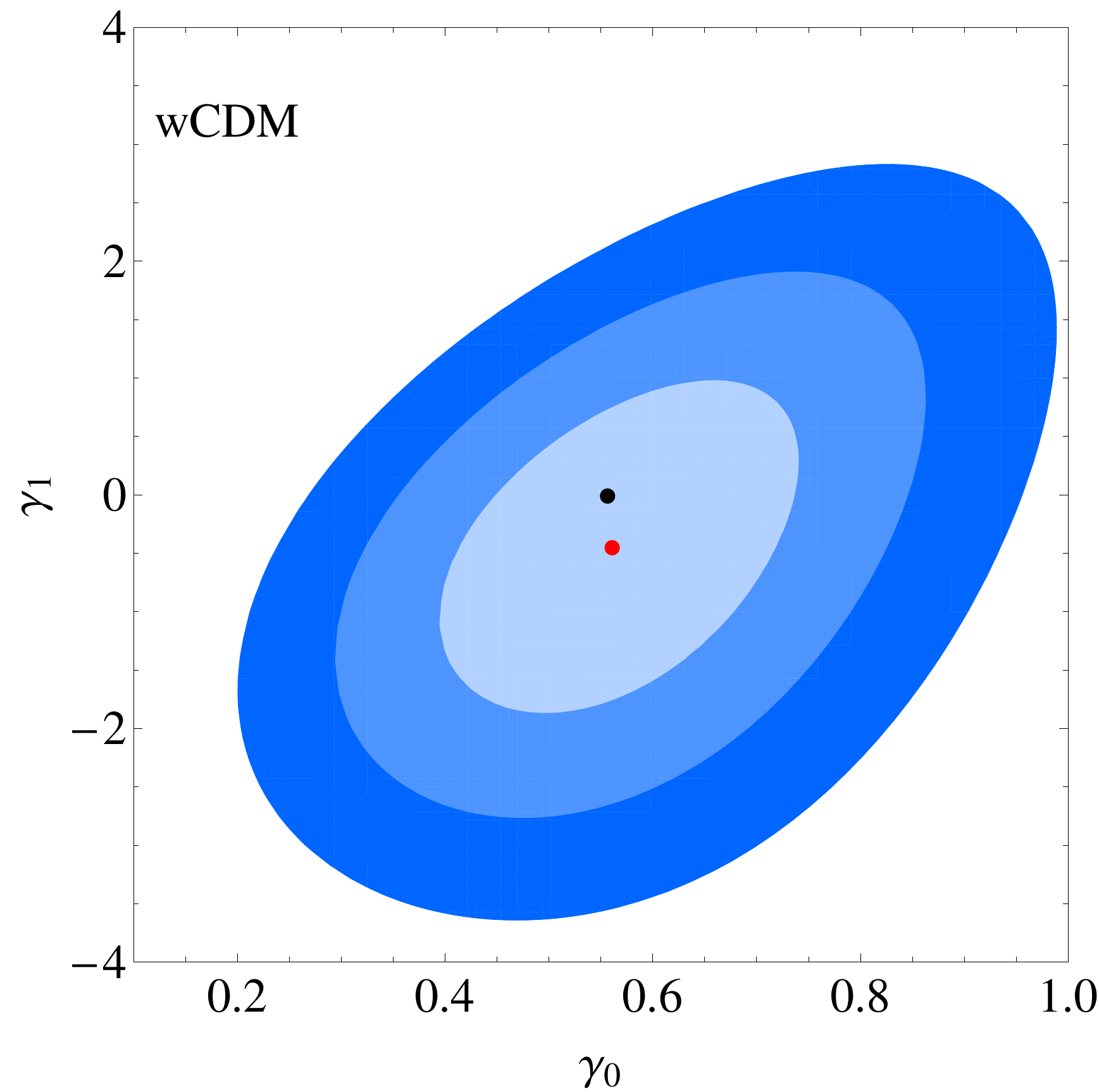}
\includegraphics[width = 0.245\textwidth]{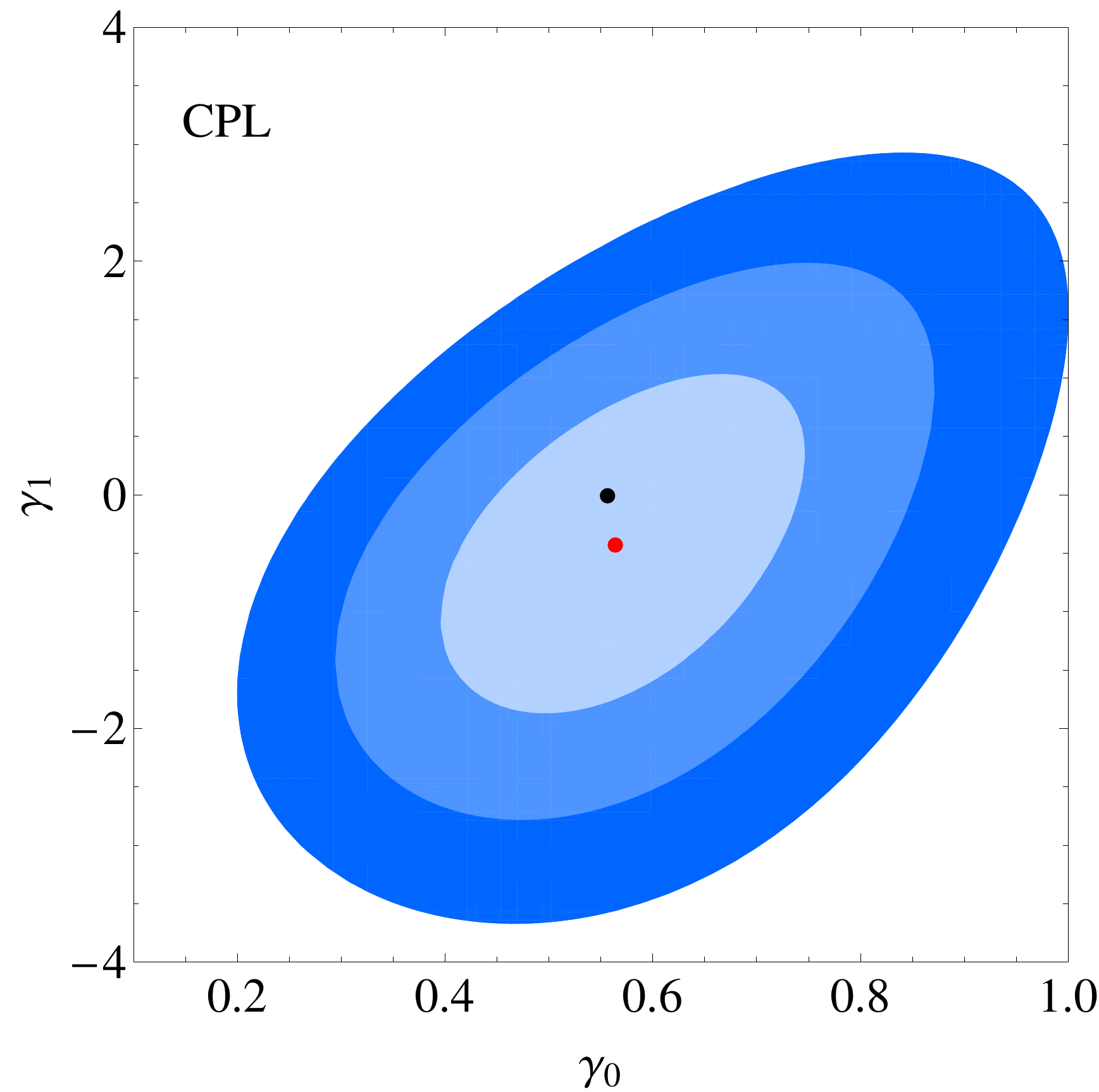}
\includegraphics[width = 0.245\textwidth]{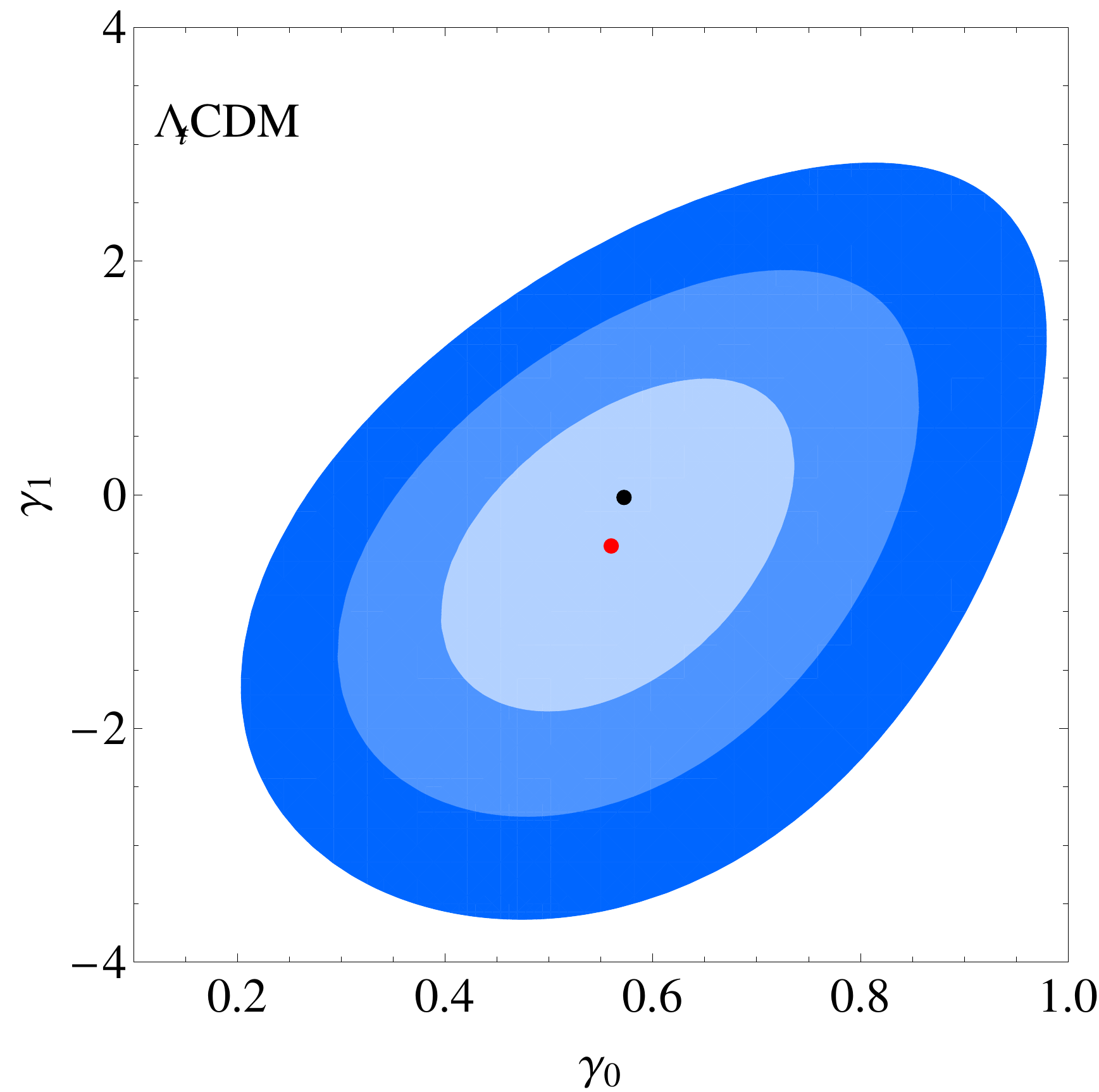}
\includegraphics[width = 0.245\textwidth]{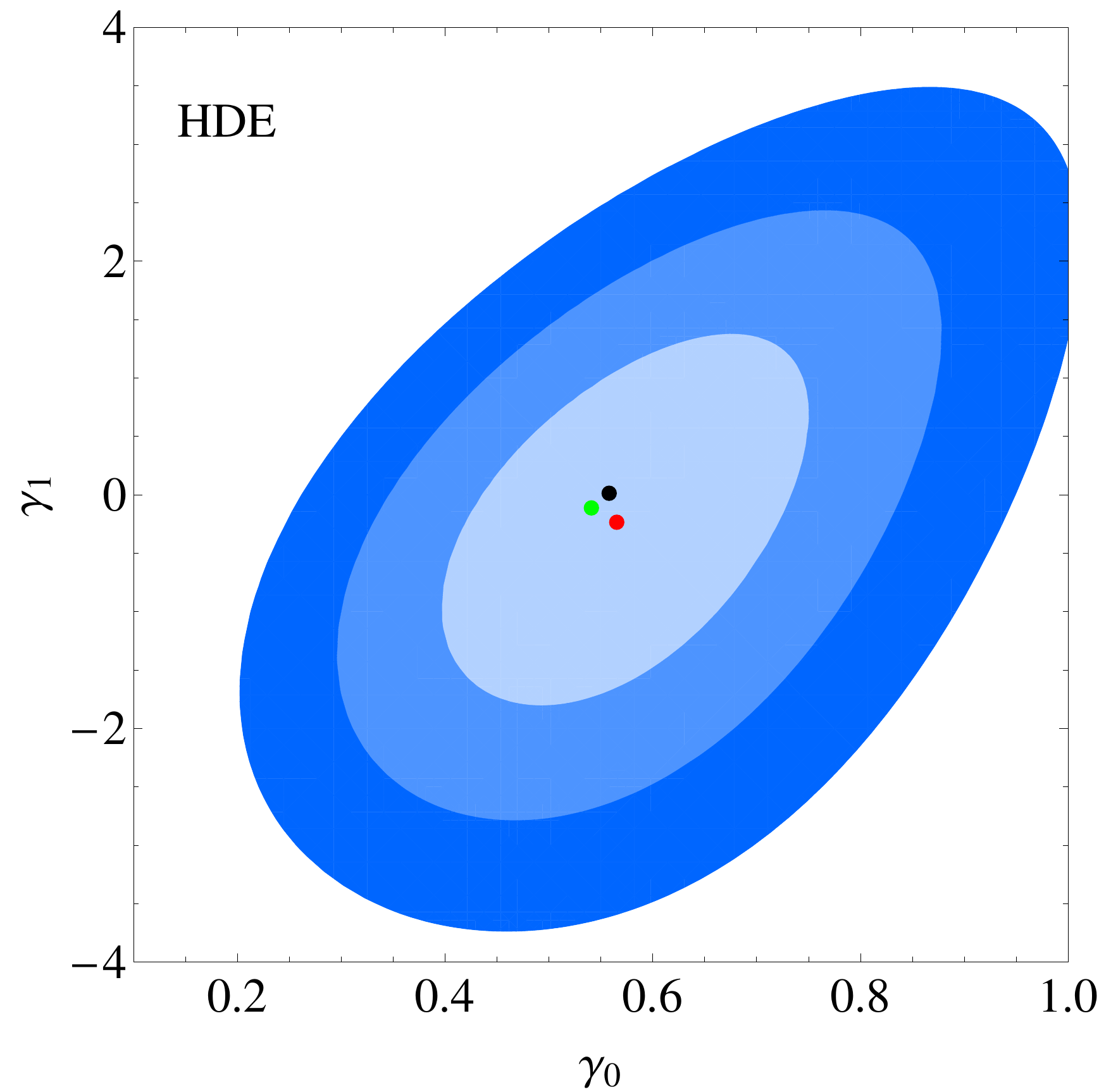}
\includegraphics[width = 0.245\textwidth]{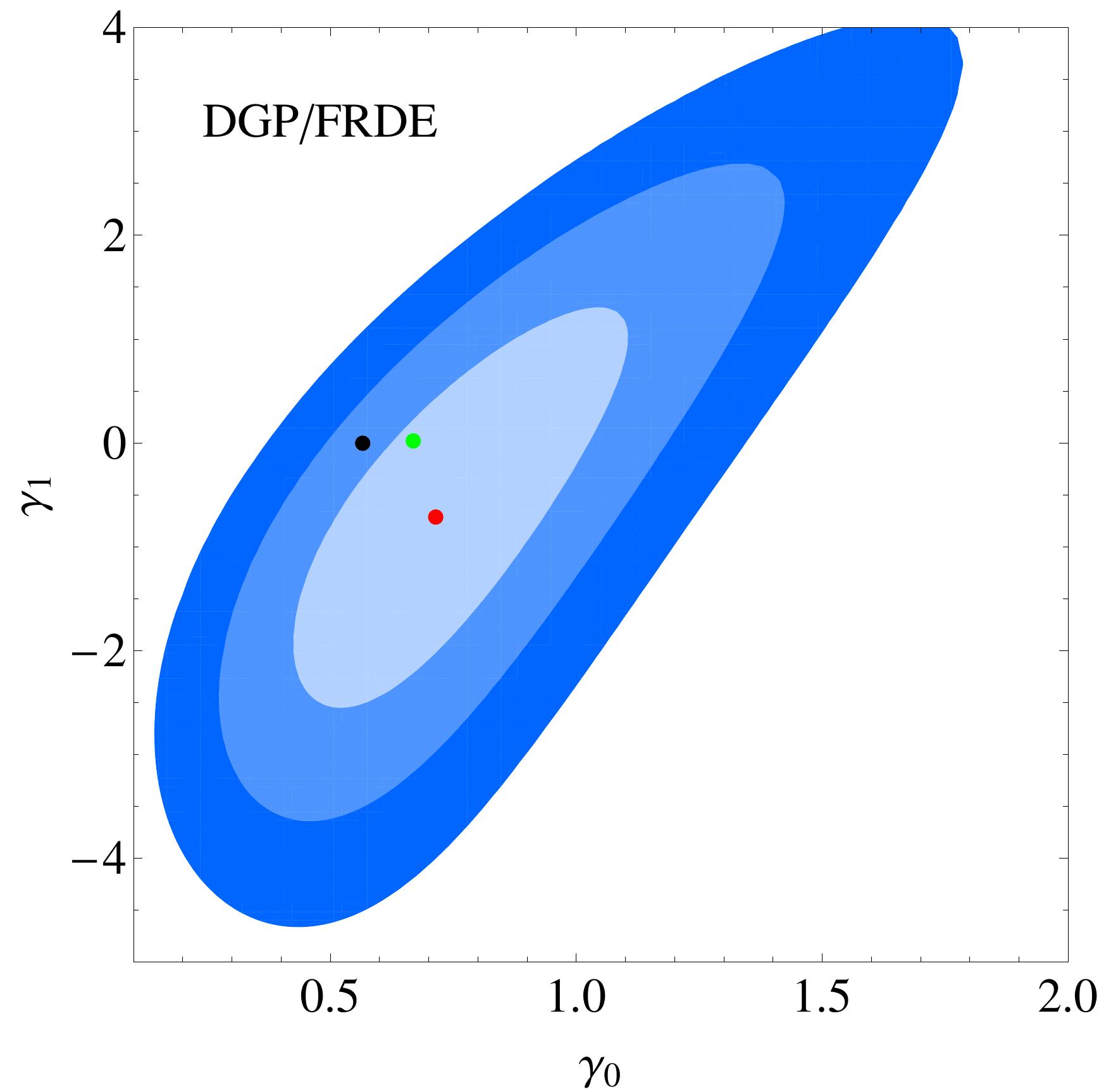}
\includegraphics[width = 0.245\textwidth]{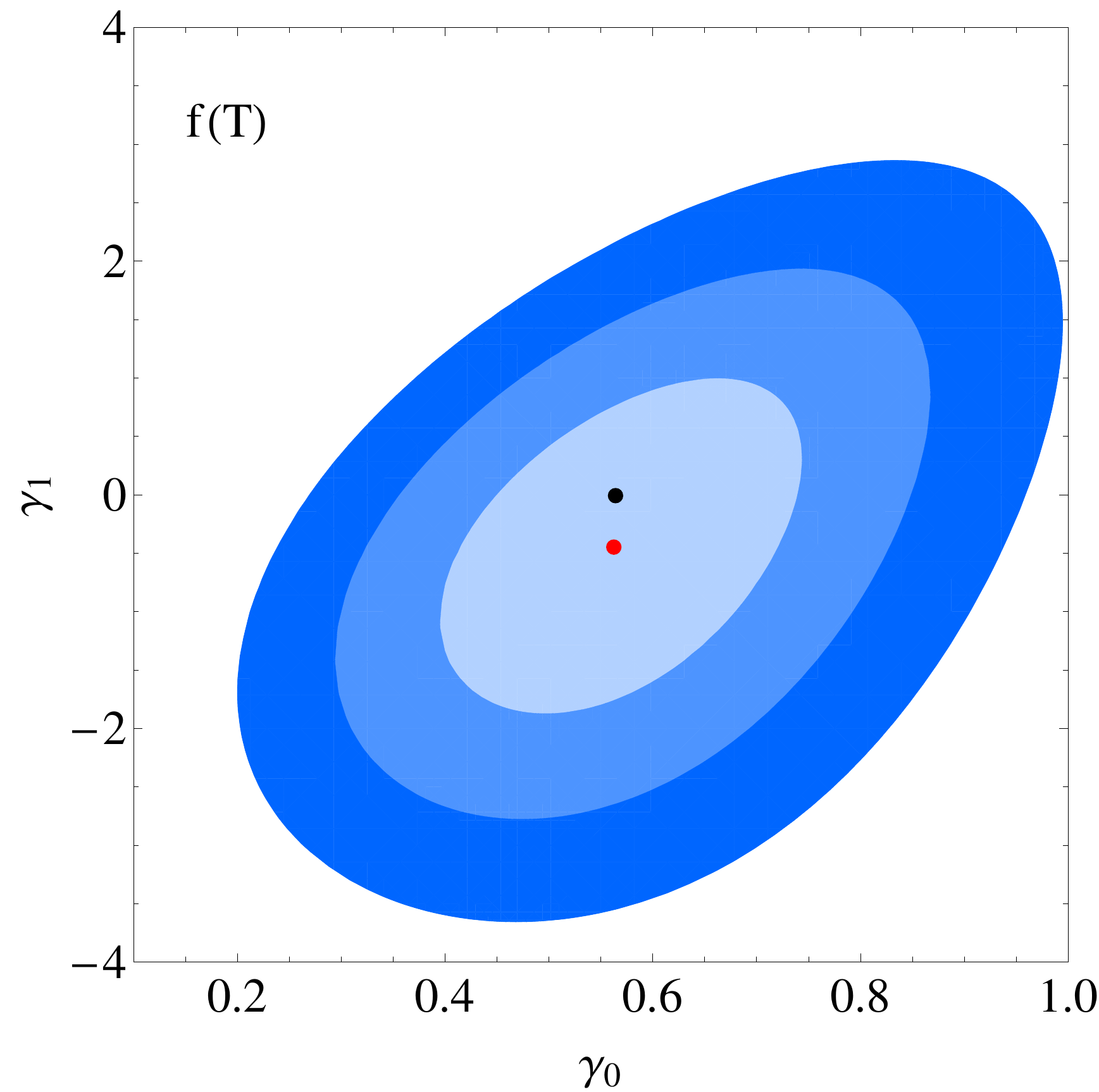}
\includegraphics[width = 0.245\textwidth]{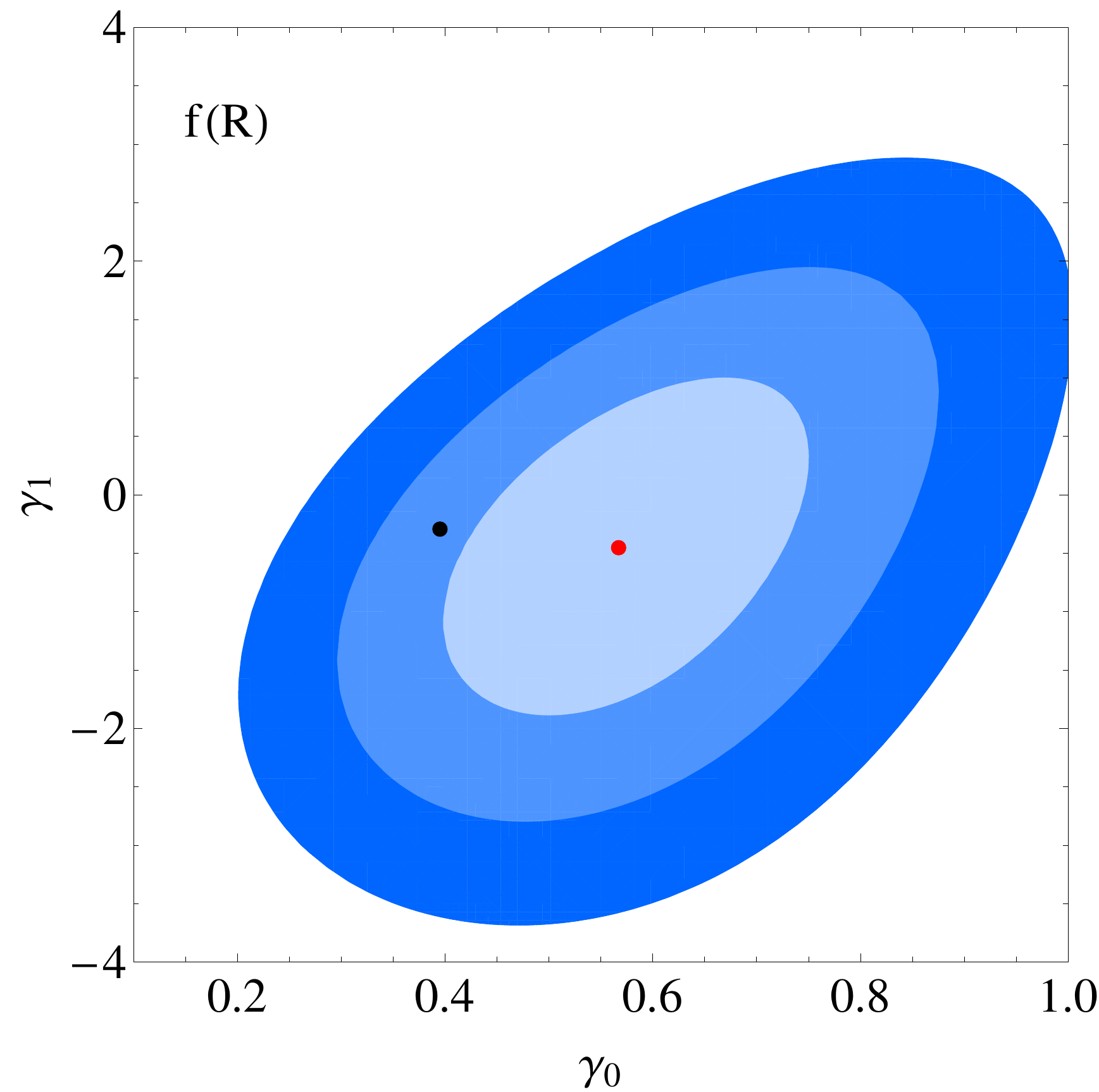}
\caption{The plots of the 1$\sigma$, 2$\sigma$ and 3$\sigma$ confidence levels in the $(\gamma_{0},\gamma_{1})$ plane with $\sigma_8$ marginalized over, for the $\Lambda$CDM, $w$CDM, CPL and $\Lambda_t$CDM models (top row) and the HDE, DGP/FRDE, $f(T)$ and $f(R)$ models (bottom row). The red dots denote the best-fit in each case, given in Table \ref{tab:growth}, while the black dots denote the theoretical predictions as given in the text. In addition, the green dots correspond to the clustered HDE model with $c_{\rm eff}^2=0$ and the DGP models.}
\label{fig:plots1}
\end{figure*}

\begin{figure*}[!t]
\centering
\includegraphics[width = 0.245\textwidth]{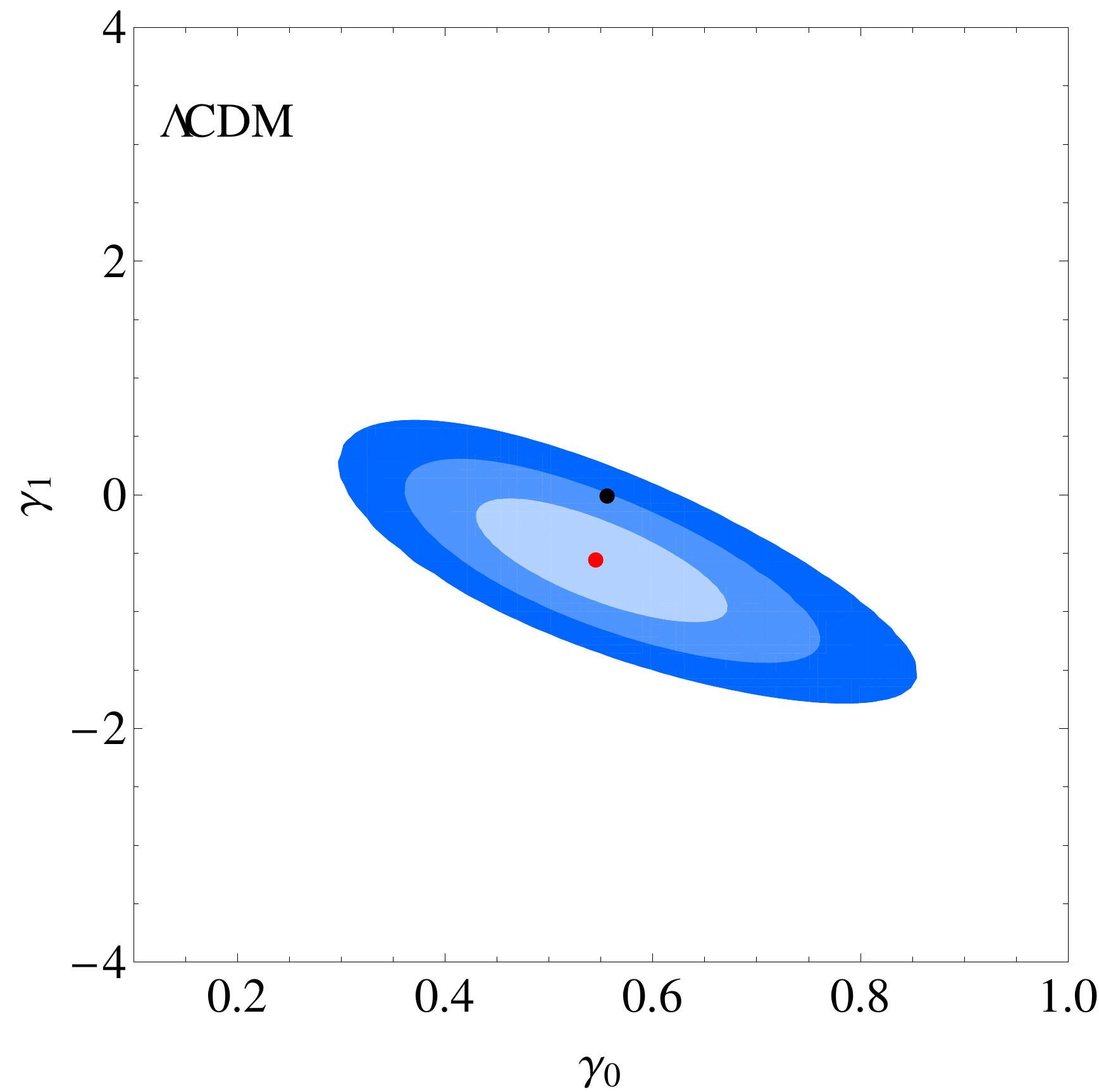}
\includegraphics[width = 0.245\textwidth]{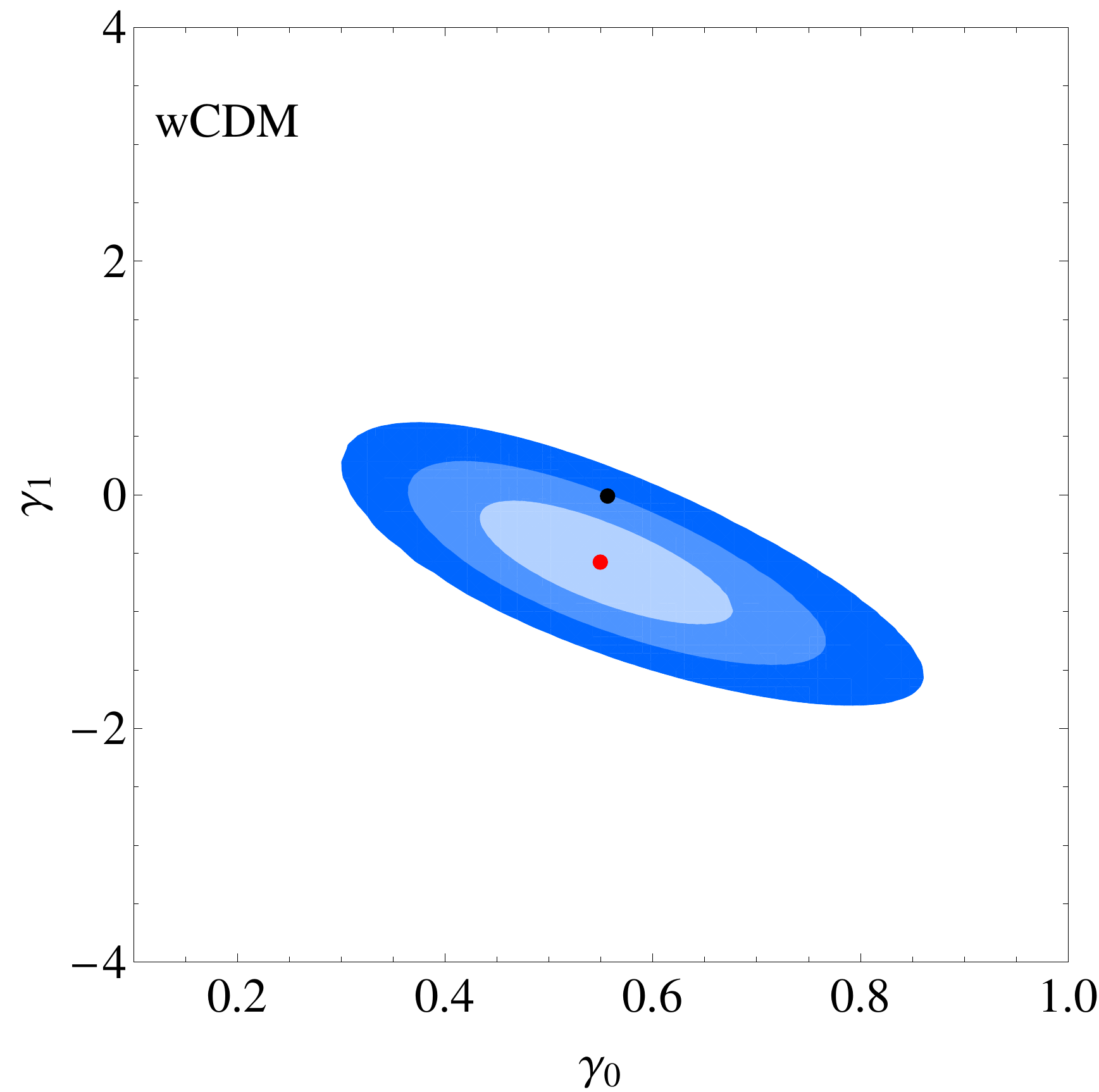}
\includegraphics[width = 0.245\textwidth]{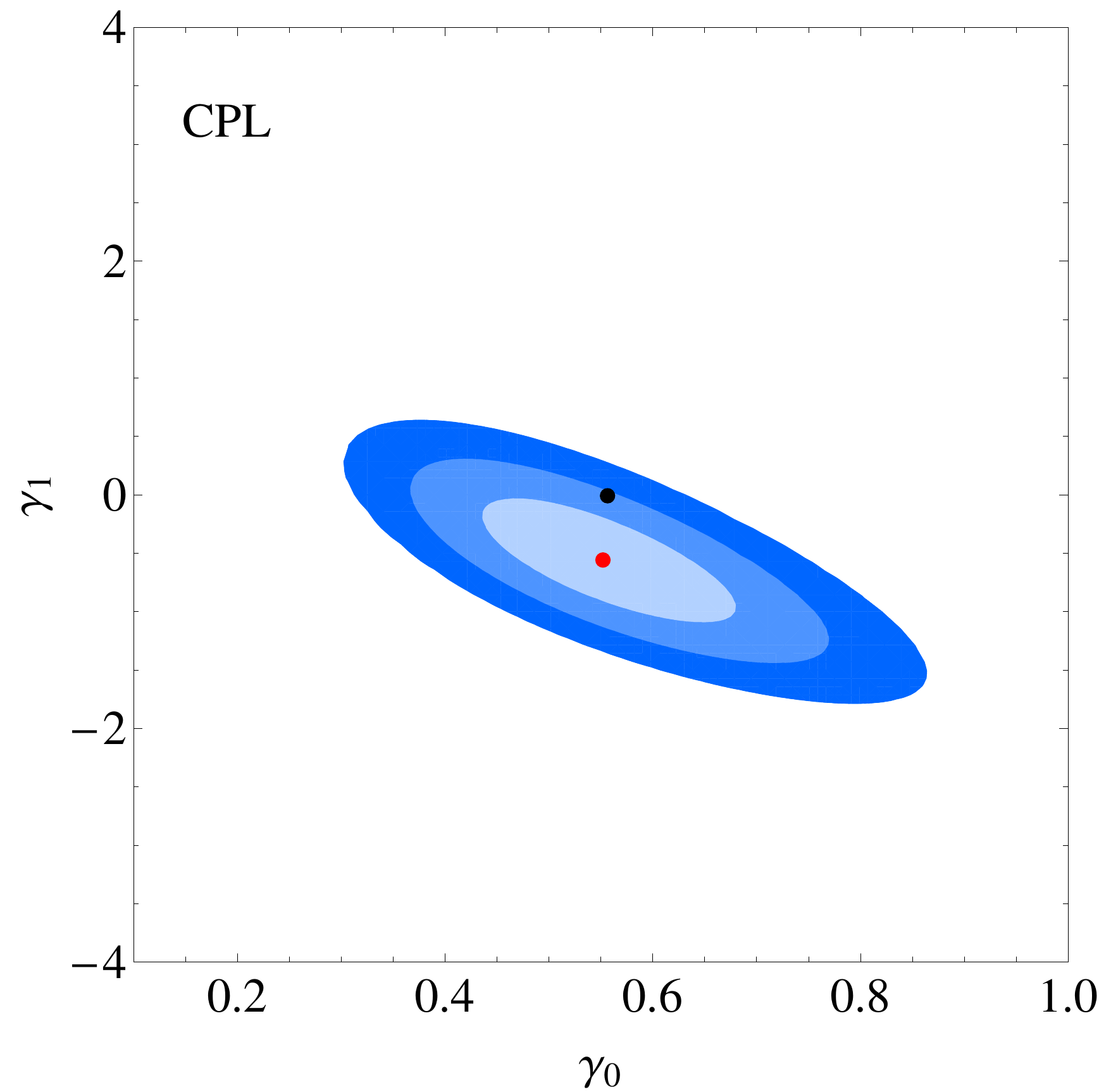}
\includegraphics[width = 0.245\textwidth]{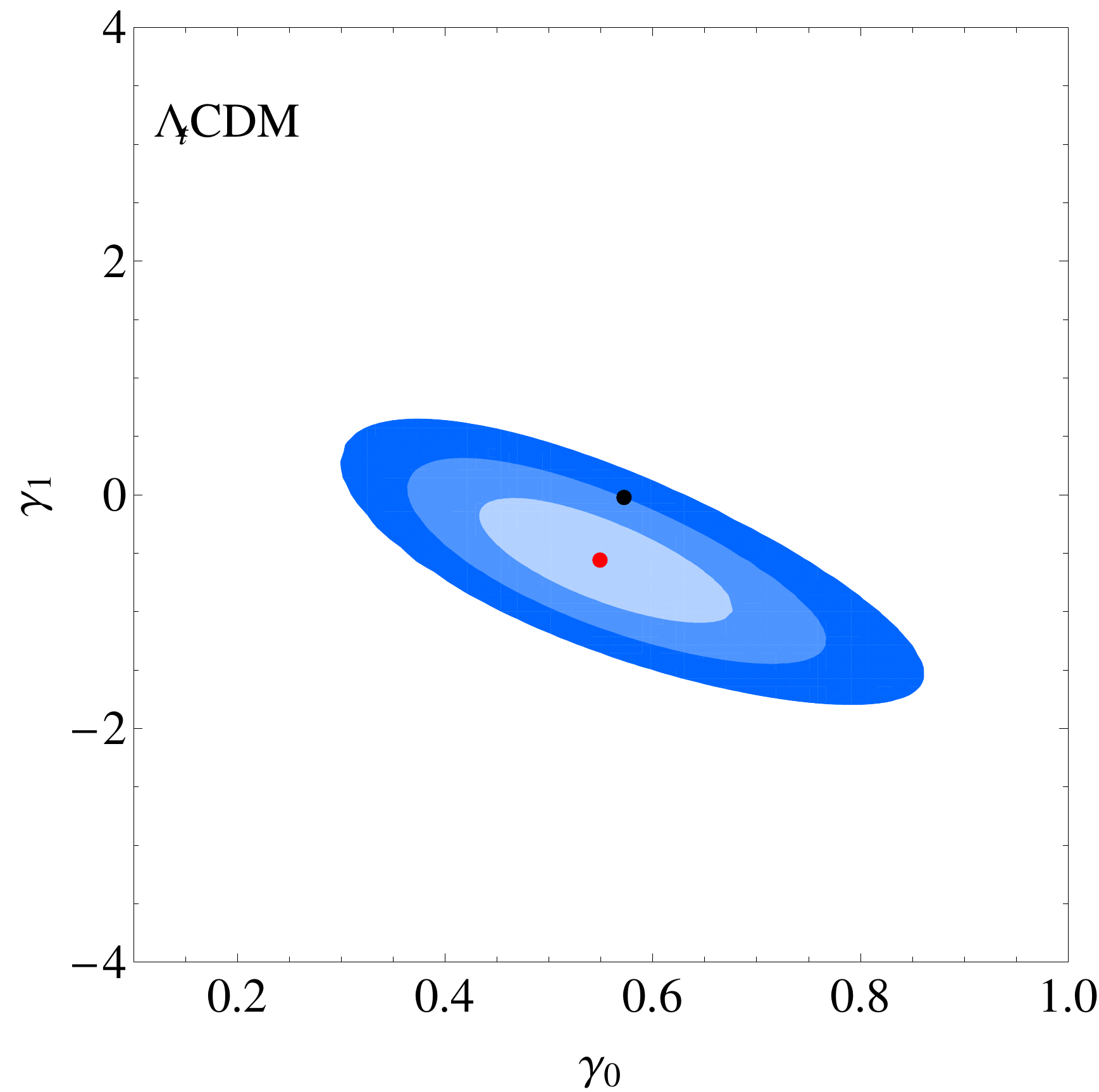}
\includegraphics[width = 0.245\textwidth]{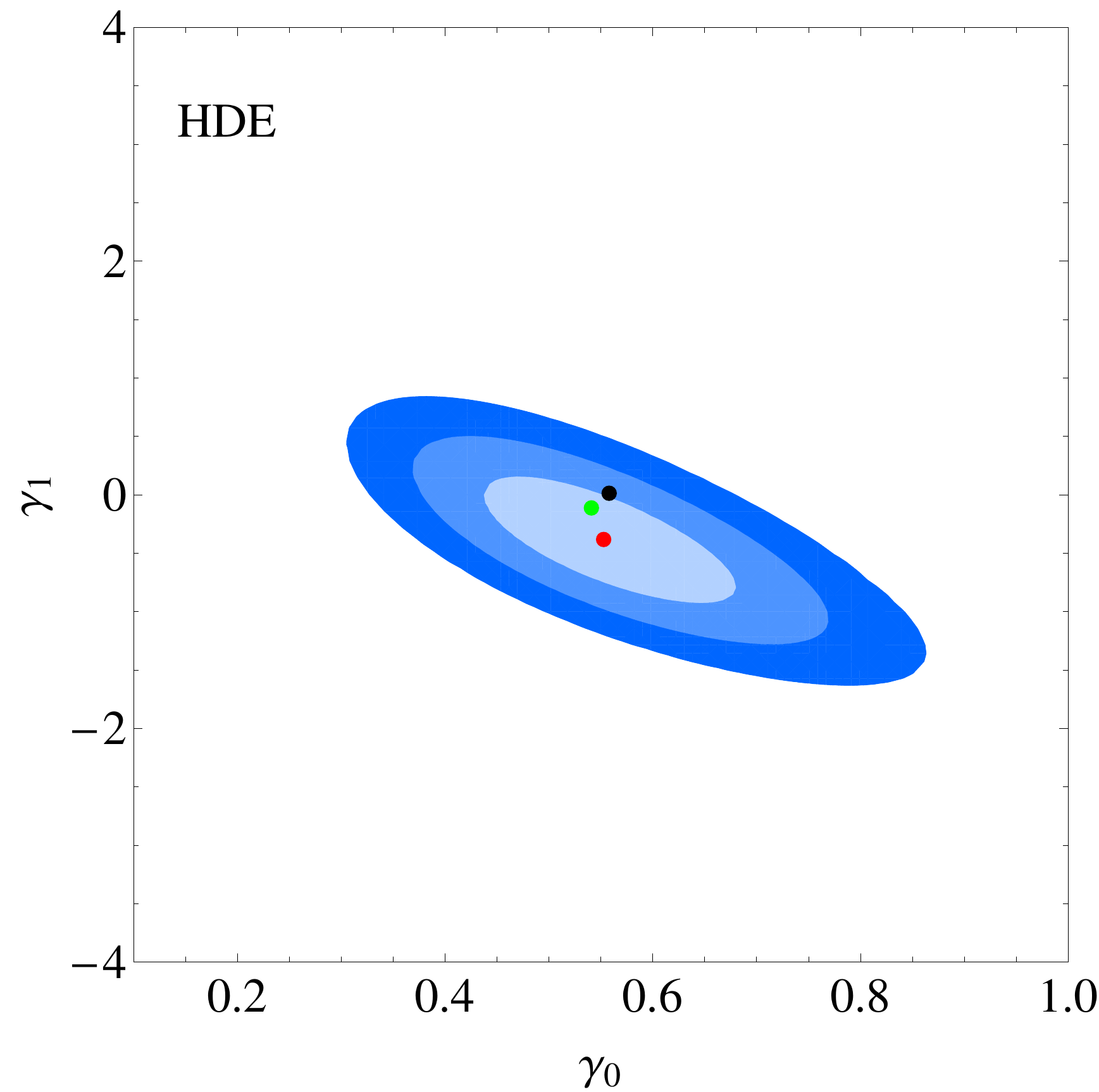}
\includegraphics[width = 0.245\textwidth]{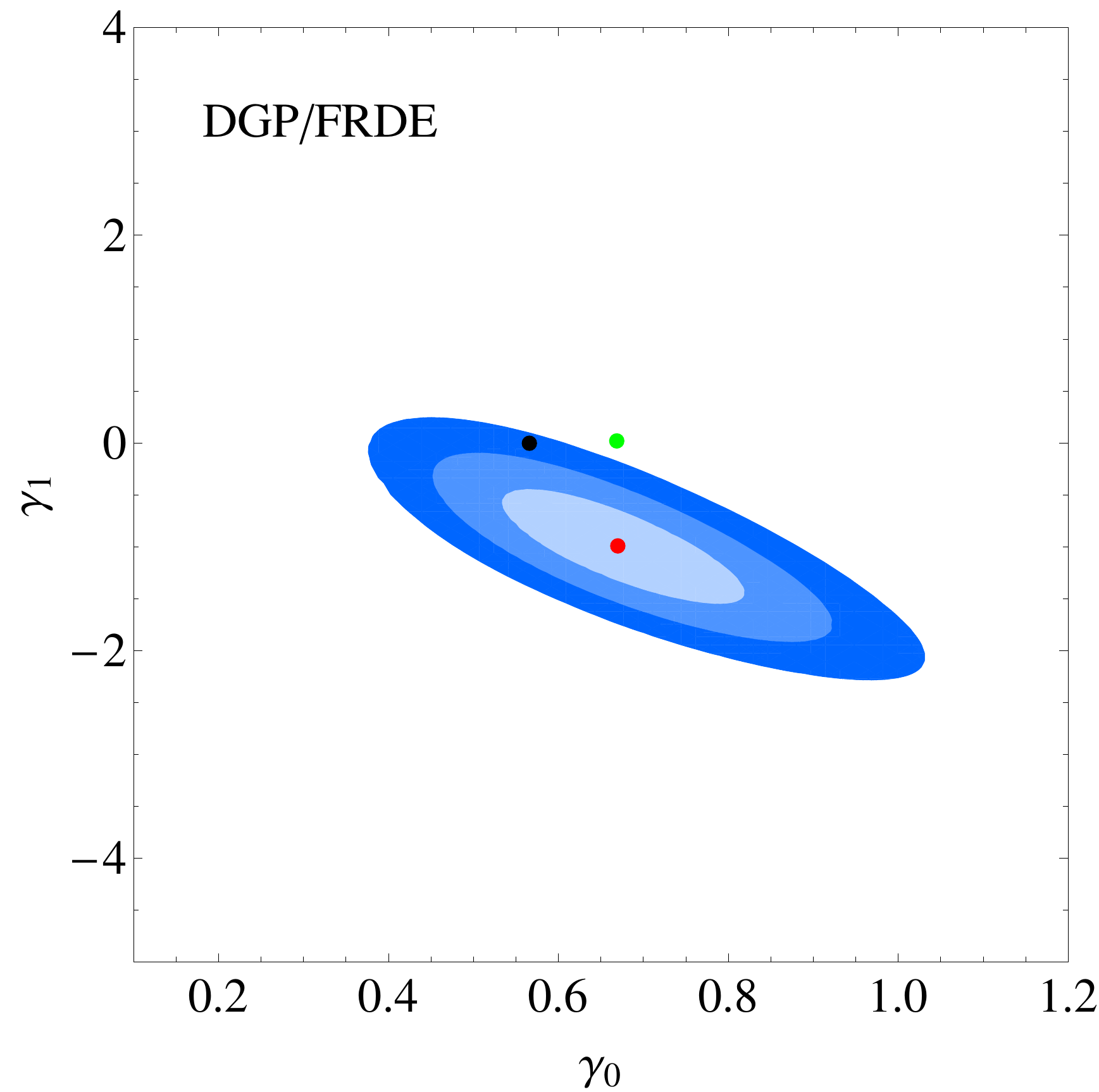}
\includegraphics[width = 0.245\textwidth]{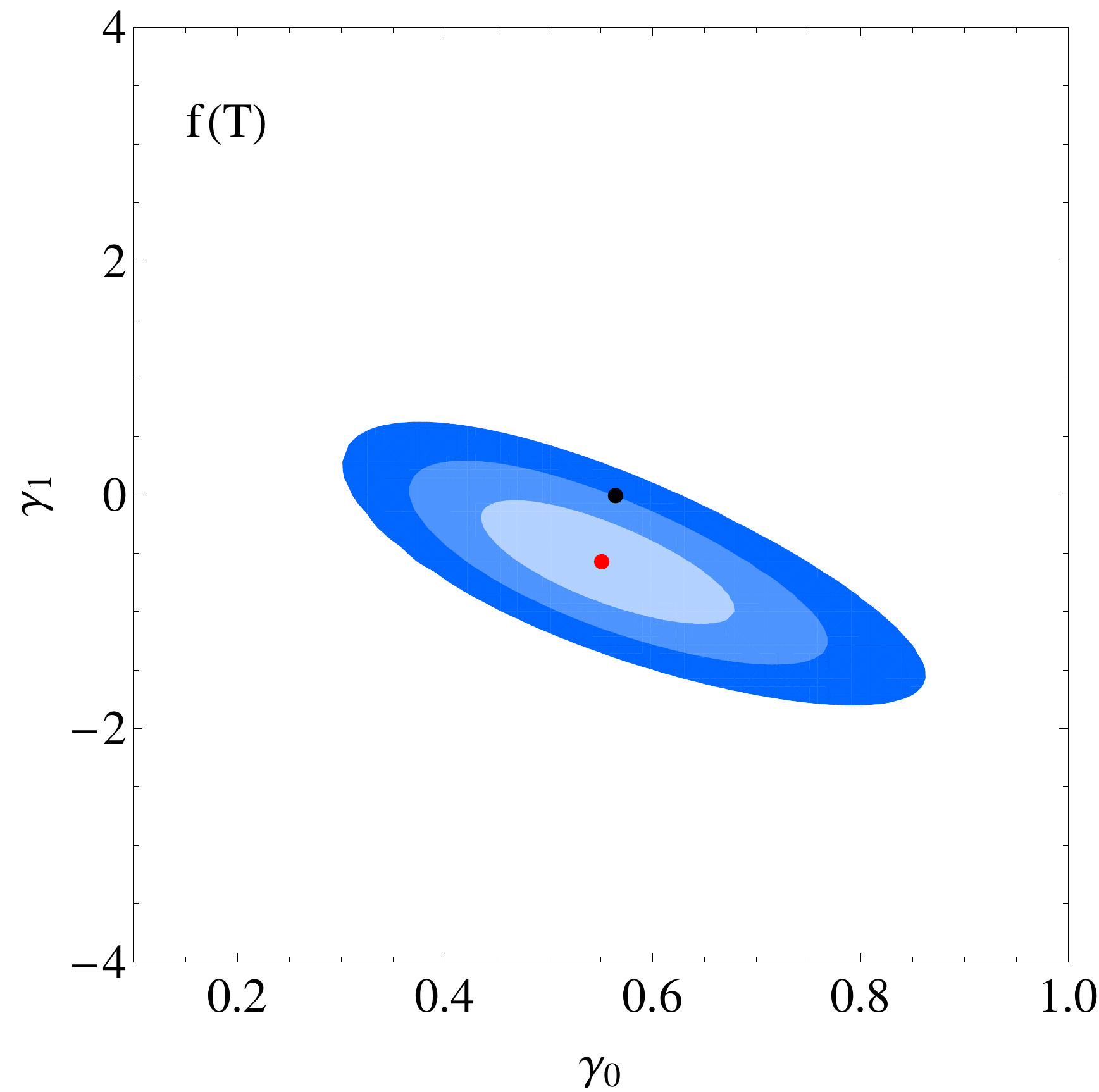}
\includegraphics[width = 0.245\textwidth]{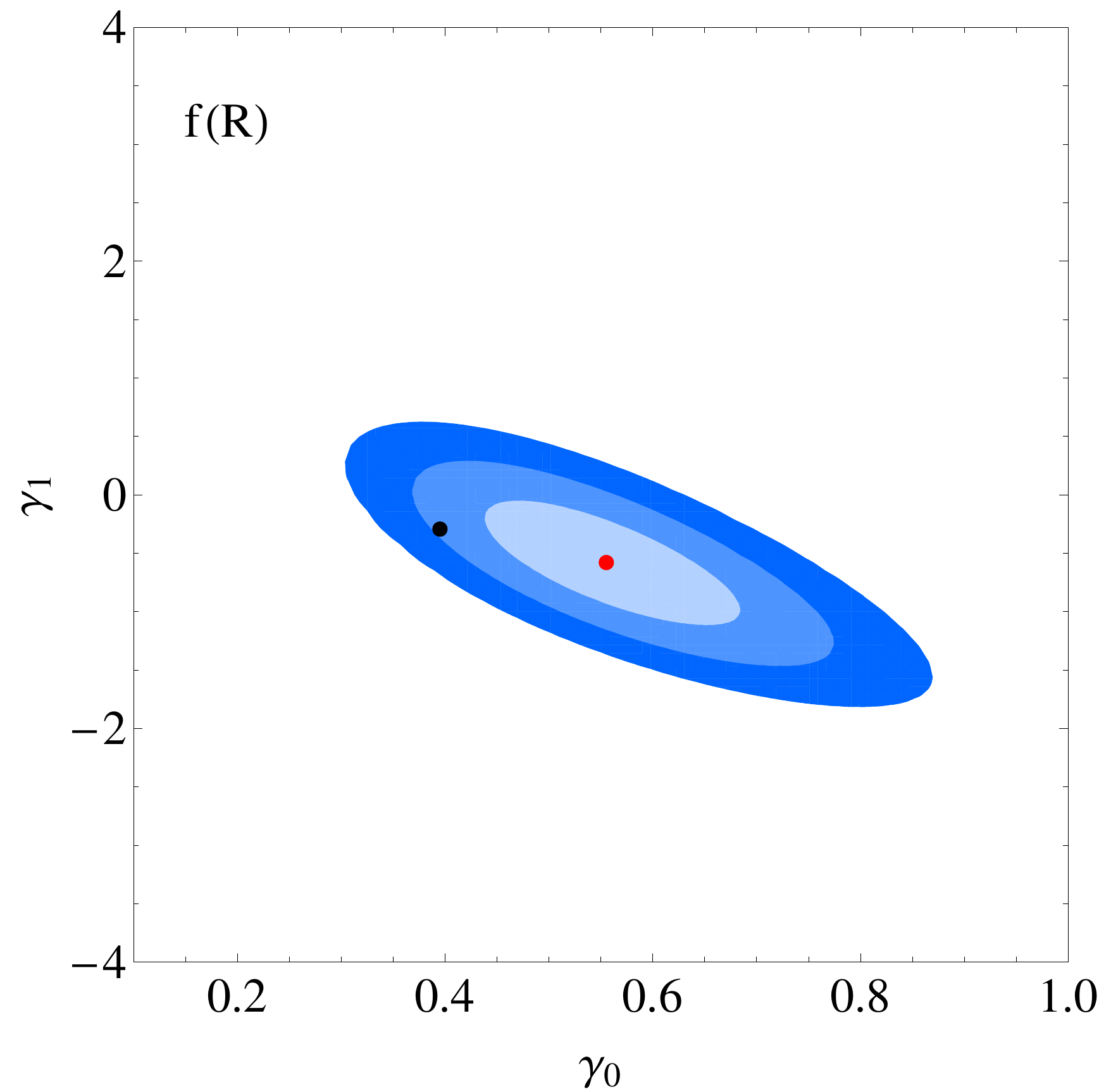}
\caption{The plots of the 1$\sigma$, 2$\sigma$ and 3$\sigma$ confidence levels in the $(\gamma_{0},\gamma_{1})$ plane with $\sigma_8$ fixed to its best-fit value, for the $\Lambda$CDM, $w$CDM, CPL and $\Lambda_t$CDM models (top row) and the HDE, DGP/FRDE, $f(T)$ and $f(R)$ models (bottom row). The red dots denote the best-fit in each case, given in Table \ref{tab:growth1}, while the black dots denote the theoretical predictions as given in the text. In addition, the green dots correspond to the clustered HDE model with $c_{\rm eff}^2=0$ and the DGP models.}
\label{fig:plots2}
\end{figure*}

In Figures \ref{fig:plots1} and \ref{fig:plots2} we present the results of our statistical analysis for the current DE cosmologies in the $(\gamma_{0},\gamma_{1})$ plane, with $\sigma_8$ marginalized over and free respectively, in which the corresponding contours are plotted for the 1$\sigma$, 2$\sigma$ and 3$\sigma$ confidence levels. On top of that we also plot the theoretical $(\gamma_{0},\gamma_{1})$ values (see section IV) of all DE models as indicated by the colored black and green dots.

Overall, we find that in the case of a marginalized $\sigma_8$, the $w$CDM, CPL, $\Lambda$CDM, $\Lambda_{t}$CDM, HDE, and $f(T)$ $(\gamma_{0},\gamma_{1})$ models are well within the $1\sigma$ borders ($\Delta \chi_{1\sigma}^{2}\simeq 2.30$; see  light blue (inner) sectors in Fig.~\ref{fig:plots1}). The rest of the DE models (DGP, FRDE and $f$CDM with $n=1$) seem to be in mild tension with the theoretical predictions for $(\gamma_0, \gamma_1)$. One the other hand, as seen in Fig.~\ref{fig:plots2} in the case of a free $\sigma_8$, we find that most models, except HDE, are in mild or in the case of the DGP in strong tension with their theoretically predicted $(\gamma_{0},\gamma_{1})$ values. Testing further the performance of our results with respect to $\sigma_{8}$
we find that there is a correlation between $\sigma_{8}$ and $\gamma_{1}$, i.e. as $\gamma_{1}$ becomes more negative $\sigma_{8}$ becomes smaller, in agreement with the results of Refs.\cite{Nesseris2013,Basilakos:2013nfa}.

Finally, we also checked that a global fit of all the free parameters, e.g. in the case of $\Lambda$CDM $\theta^{i}=(\alpha, \beta,\Omega_{m0},\Omega_{b0}h^2, h,\gamma_0, \gamma_1,\sigma_8)$, gives exactly the same fit as our two step process. This confirms our assumption that the background parameters $(\alpha, \beta,\Omega_{m0},\Omega_{b0}h^2, h)$ and perturbation order parameters $(\gamma_0, \gamma_1,\sigma_8)$ are uniquely fixed by their corresponding data.

\subsection{Analysis with mock data}
In this section we briefly discuss forecasts of our methodology with mock $f\sigma_8$ data based on a $\Lambda$CDM cosmology with $(\Omega_{m0},\sigma_8)=(0.3,0.8)$. The mocks were created following the methodology of Ref.~\cite{Nesseris:2014qca} having in mind a setup similar to Euclid-like and LSST-like surveys. To actually create the data, we evaluated the $f\sigma_8(z)$ for the $\Lambda$CDM cosmology, uniformly distributed in the range $z \in [0, 2]$ divided into $10$ equally spaced bins of step d$z=0.2$. The $f\sigma_8(z_i)$ at each point was estimated by adding its theoretical value to a Gaussian error and assigning an error of $1\%$ of its value, which is in agreement with the expected LSST accuracy as described in Refs.~\cite{Huterer:2013xky,Abell:2009aa}.

\begin{figure*}[!t]
\centering
\includegraphics[width = 0.49\textwidth]{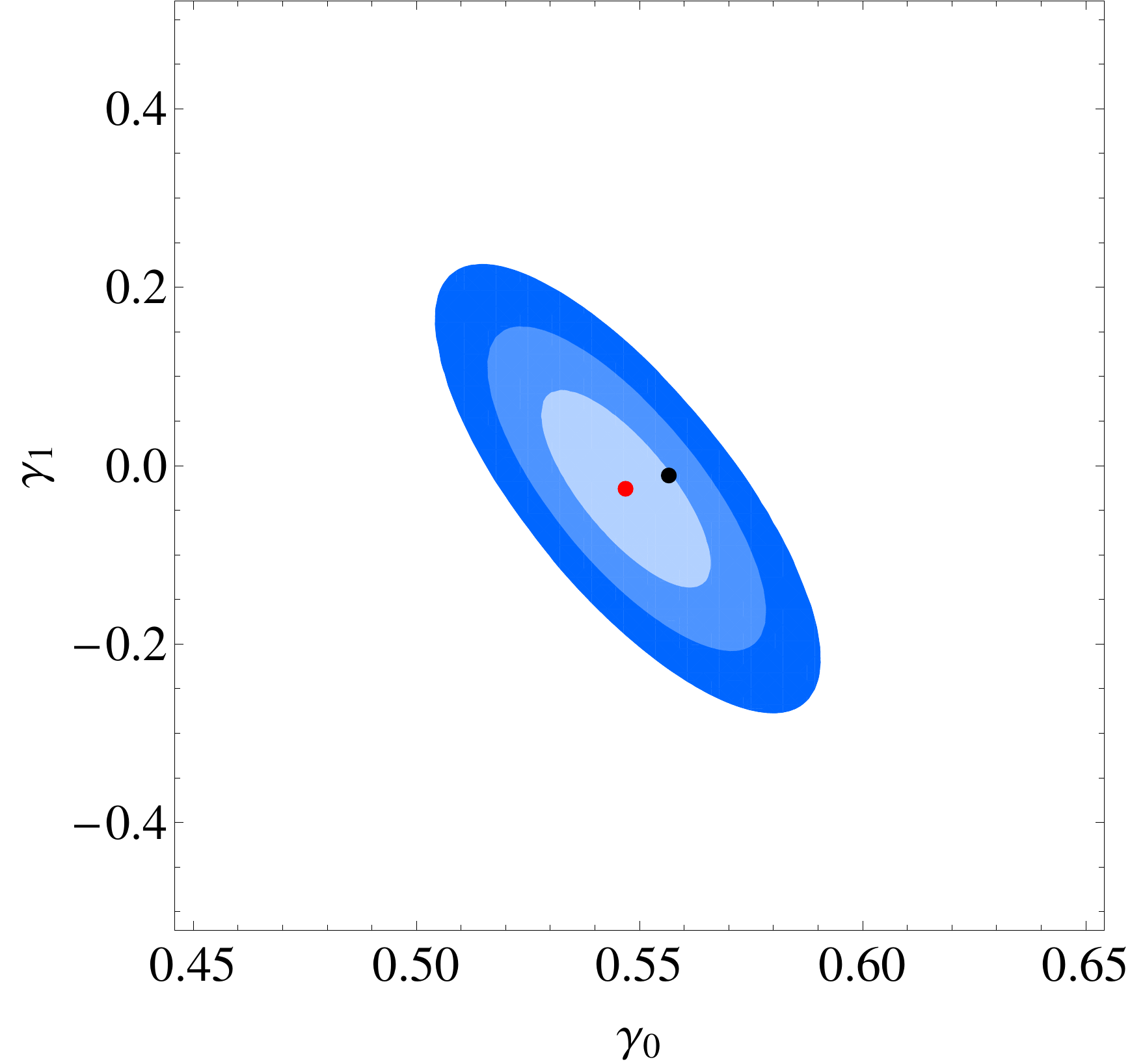}
\includegraphics[width = 0.49\textwidth]{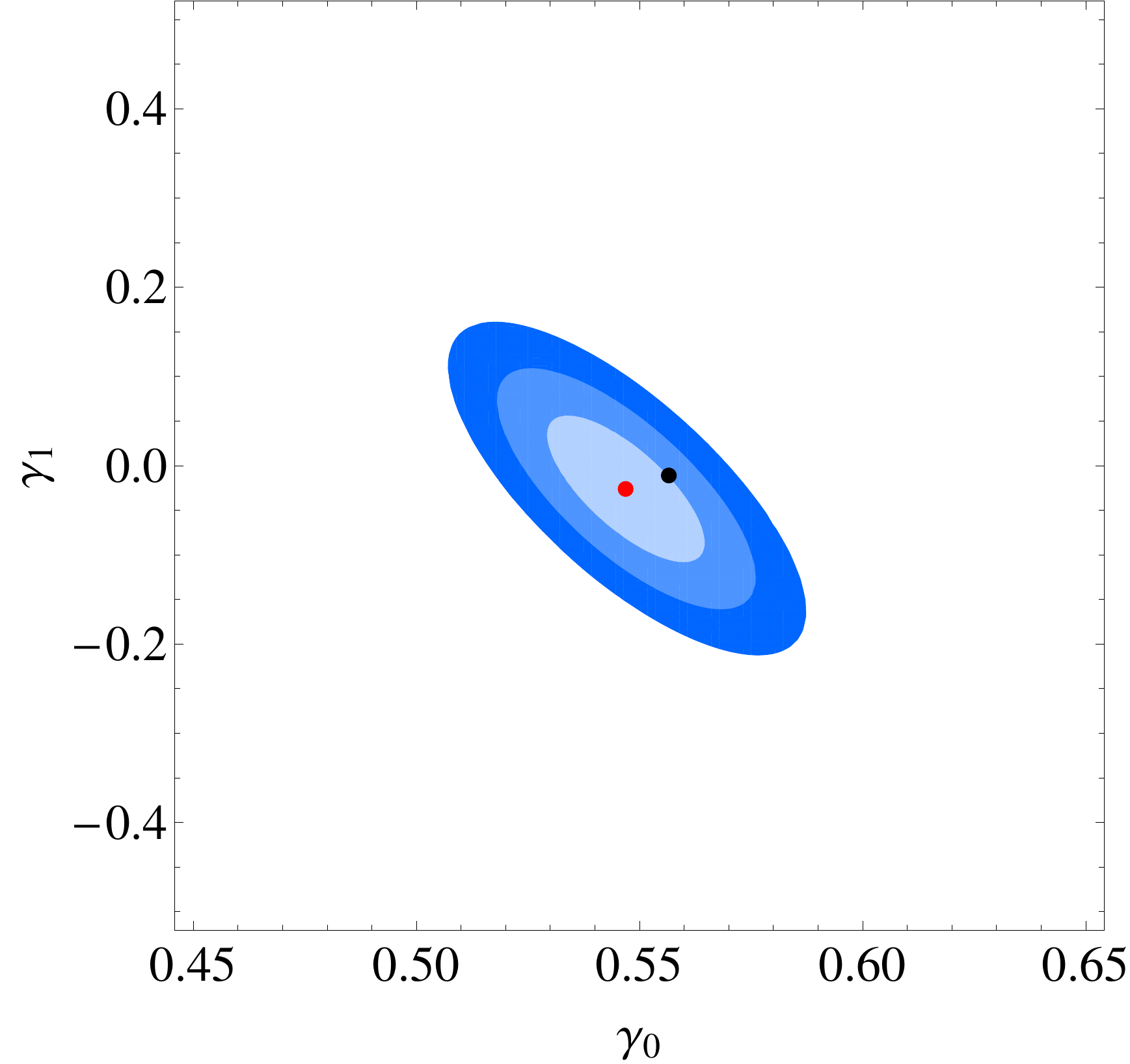}
\caption{The plots of the 1$\sigma$, 2$\sigma$ and 3$\sigma$ confidence levels in the $(\gamma_{0},\gamma_{1})$ plane with $\sigma_8$ either marginalized over (left) or fixed to its best-fit value (right) for the $\Lambda$CDM  model and the mock LSST-like data. The red dots denote the best-fit in each case, given in the text, while the black dots denote the theoretical predictions as given in the text.}
\label{fig:plots3}
\end{figure*}

Following the same analysis as before, we consider the same two strategies for dealing with $\sigma_8$, i.e. first by marginalizing over it and second by fitting it along with the other two parameters. In the first case we find $\gamma_0=0.547\pm0.019$ and $\gamma_1=-0.026\pm0.111$, while in the second one we have $\gamma_0=0.547\pm0.018$, $\gamma_1=-0.026\pm0.111$ and $\sigma_8=0.802\pm0.004$. In Fig.~\ref{fig:plots3} we present the results of our statistical analysis for the mock $f\sigma_8$ data in the $(\gamma_{0},\gamma_{1})$ plane, with $\sigma_8$ marginalized over (left panel) and free (right panel) respectively, in which the corresponding contours are plotted for the 1$\sigma$, 2$\sigma$ and 3$\sigma$ confidence levels.

As can be seen, the constraints on $(\gamma_{0},\gamma_{1})$ are much more in line with the theoretical predictions (black dots in Fig.~\ref{fig:plots3}), thus making it possible to discriminate between DE models and proving that our methodology will be extremely useful with the upcoming data in the near future.

\section{Conclusions}
\label{conclusions}
We studied the growth index beyond the concordance $\Lambda$CDM model by utilizing several forms for the dark energy. In the first part of our article, we implemented an overall likelihood analysis using the most recent high quality cosmological data (SNIa, CMB shift parameter and BAOs), thereby putting tight constraints on the main cosmological free parameters. At the level of the resulting Hubble function, we showed that the majority of dark energy models (apart from HDE, DGP and Finsler-Randers cosmologies), are statistically indistinguishable (within 1$\sigma$) from a flat $\Lambda$CDM model, as long as they are confronted with the above background geometrical data. Of course, the DGP can readily be ruled out as it has a $\delta \chi^2\sim 90$ from $\Lambda$CDM.

At the perturbation level, not only do the aforementioned DE models reproduce the $\Lambda$CDM Hubble expansion, but we also found that by using their
$\chi^2_{\rm min}$ and AIC values, all models fit equally well the growth rate data and are statistically indistinguishable from the $\Lambda$CDM model on the basis of their growth index evolution.

However, it should be noted that in the case of a free $\sigma_8$ parameter the best fit values of $(\gamma_0,\gamma_1)$ for most models, except HDE, are in mild to strong tension with their corresponding theoretically predicted values, see Fig.~\ref{fig:plots2}. On the contrary, this is not so apparent in the case of a marginalized $\sigma_8$, as the contours are now larger due to the marginalization. In this case, after inspection of the contours of Fig.~\ref{fig:plots1} our results can be summarized in the following statements (for more details see section V):

\begin{itemize}
\item Three models, ie., DGP, FRDE and $f$CDM can be distinguished since they are in tension with the growth data and they show strong and significant variations with respect to the concordance $\Lambda$ model.

\item Four DE models, namely, $w$CDM, CPL, $\Lambda_{t}$CDM, HDE and $f(T)$ are in agreement with the growth data and they cannot be distinguished from the $\Lambda$CDM model at any significant level.
\\
\end{itemize}

The reason we considered both approaches, i.e. marginalizing over and fitting $\sigma_8$, is that as this is a derived parameter, the uncertainty in its measurement is still rather high, thus affecting the interpretation of our analysis. Furthermore, we found that the growth rate data consistently and for all models, prefer a rather low value for $\sigma_8$ of approximately $\sigma_8\simeq 0.687\pm0.095$, which is in significant tension with the Planck result of $\sigma^{Planck}_8= 0.831\pm0.013$, see \cite{Ade:2015xua}. This difference corresponds to a $1.5\sigma$ discrepancy from the growth rate data point of view or $9.5\sigma$ from the Planck data point of view. It should be noted that the observed suppression of power of the late Universe observables, e.g. low $\sigma_8$ and the chronic tension between the CMB and low-redshift observables has already been discussed in the literature, see e.g. Ref.~\cite{Kunz:2015oqa}.

To conclude, the main benefit of our analysis is that even though at a first glance all of the models seem indistinguishable at the statistical level given their $\chi^2_{\rm min}$ and AIC values, see Tables \ref{tab:growth} and \ref{tab:growth1}, when compared with their theoretically predicted $(\gamma_0,\gamma_1)$ ones, we can see there is a significant inconsistency. Since the models are known to be internally consistent this means either that there is a problem with the growth-rate data or that there is new physics emerging at low redshifts. Also, by using mock growth rate data we demonstrated that this will be made more clear with future dynamical data (based mainly on {\em LSST}), which are expected to improve significantly the relevant constraints (especially on $\gamma_1$ and $\sigma_8$) and thus possibly resolve the issue with the lack of power at low redshifts discussed in Ref.~\cite{Kunz:2015oqa}, but also shedding light on the nature of dark energy on cosmological scales.

\section*{Acknowledgements}
The authors would like to thank W.~Cardona for pointing out a typo in the text.

S.B. acknowledges support by the Research Center for Astronomy of the Academy
of Athens in the context of the program ``{\it Tracing the Cosmic Acceleration}''.

S.N. acknowledges support by the Research Project of the Spanish MINECO, Grant No. FPA2013-47986-03-3P, the Centro de Excelencia Severo Ochoa Program Grant No. SEV-2012-0249 and the Ram\'{o}n y Cajal program through Grant No. RYC-2014-15843.

\end{document}